\documentclass[12pt]{article}
\usepackage{amssymb}
\usepackage{amsmath}
\usepackage{cancel}
\usepackage{slashed}
\usepackage{fullpage}
\usepackage{color}
\usepackage{braket}
\usepackage{graphicx}
\usepackage[small]{subfigure}
\usepackage{multirow}
\usepackage{verbatim}
\usepackage{mathtools} 
\usepackage{amsthm}
\usepackage{simplewick}
\usepackage{cite}

\renewcommand{\neq}{\slashed{=}}

\def\cD{\mathcal{D}}

\def\cL{\mathcal{L}}
\def\cM{\mathcal{M}}

\def\cO{\mathcal{O}}
\def\cP{\mathcal{P}}

\def\cW{\mathcal{W}}

\def\cY{\mathcal{Y}}

\def\And{\quad {\rm and} \quad}
\def\Or{\quad {\rm or} \quad}
\def\For{\quad {\rm for} \quad}

\def\dg{\dagger}

\newcommand{\cn}[1]{\cancel{#1}}
\newcommand{\Sl}[1]{\slashed{#1}}
\newcommand{\fd}[2]{\parbox{#1}{\includegraphics[width=#1]{#2}}}

\def\be{\begin{equation}}
\def\ee{\end{equation}}

\def\l{\langle}
\def\r{\rangle}
\def\a{\alpha}
\def\ad{{\dot\alpha}}
\def\b{\beta}
\def\bd{{\dot\beta}}

\newcommand{\Eq}[1]{Eq.~\eqref{#1}}

\newcommand{\LQCD}{{\cL_{\text{QCD}}}}
\newcommand{\LEFT}{{\cL_{\text{EFT}}}}
\def\LPeq{\cong}

\newcommand{\kperm}{ { \{k\} \hookrightarrow \{ \ell_i \}  } }

\newcommand{\pie}{}

\title{An on-shell approach to factorization}
\author{Ilya Feige and  Matthew D. Schwartz}

\begin{document} 
\maketitle


\begin{abstract}
Factorization is possible due to the universal behavior of Yang-Mills
theories in soft and collinear limits. 
Here, we take a small step towards a more transparent understanding
of these limits by proving a form of perturbative factorization  at tree-level using on-shell spinor helicity methods. 
We present a concrete and self-contained expression of factorization
in which matrix elements in QCD are
related to products of other matrix elements in QCD up to leading order in a power-counting parameter determined by the momenta
of certain physical on-shell states.
Our approach uses only the scaling of momenta in soft and collinear limits, avoiding
any assignment of scaling behavior to unphysical (and gauge-dependent) fields. 
The proof of factorization exploits many advantages of helicity spinors, such as the freedom
to choose different reference vectors for polarizations in different collinear sectors. 
An advantage of this approach is that once factorization is shown to hold in QCD, the transition to Soft-Collinear Effective Theory is effortless.
\end{abstract}

\newpage


\section{Introduction}
\label{sec:intro}

That perturbative calculations in a strongly coupled theory like quantum chromodynamics (QCD) 
can ever be related to experimental data is
due to two remarkable properties:
asymptotic freedom and factorization. Of these, asymptotic freedom is much better understood. Indeed,
the asymptotic behavior of a theory  can usually be established from
the ultraviolet divergences in 1-loop amplitudes. It is a short distance property. Factorization, on the other hand, is a long-distance property. In its most intuitive, and most useful form, factorization states that cross sections in QCD can be calculated up to power corrections in
some small scale $\lambda$
by convolutions of universal (and often nonperturbative) objects, such as parton-distribution functions, and perturbative, but process-dependent matrix elements.

There are both nonperturbative and pertubative aspects to factorization. 
On the nonperturbative side, one would ideally like to prove that factorization holds up to corrections in $\lambda=m_P/Q$ with $m_P$ the proton (or some other particle) mass and $Q\gg m_P$ some high-energy scale. Unfortunately, to have access to $m_P$, which is a non-perturbative quantity in QCD, one needs access to nonperturbative physics. Instead, most approaches to factorization simply assume that operator matrix elements in hadronic states, and final-state hadronization effects,
do not violate naive scaling expectations. Then they use perturbation theory and scaling arguments to relate those matrix elements among different processes. The most familiar example of this approach is the universality of the parton distribution functions (PDFs). Factorization implies (or should imply, if it were proven generally) that the same operator matrix elements representing the PDFs appear in the calculation of a great variety of physical processes. Since a preponderance of experimental evidence confirms this universality, providing a general proof seems almost academic.

On the other hand, there are purely perturbative aspects to factorization with great practical importance.
For example, factorization is a first step in performing resummation which is necessary to reproduce even qualitative
features of certain distributions. Classic examples are event shapes, particularly at $e^+e^-$ colliders~\cite{Catani:1992ua,Schwartz:2007ib,Becher:2008cf,Chien:2010kc,Becher:2012qc,Chiu:2012ir}, and processes with hard well-separated objects at hadron colliders (such as photon plus jet production). In these cases, having a precise statement of factorization, with operator definitions of all the objects involved, allows one to compute distributions to all orders in $\alpha_s$ in certain singular regions, or to provide approximate fixed-order calculations for inclusive cross sections.
Since exact calculations in QCD beyond next-to-leading order can be extraordinarily challenging, having an alternative approach to produce numerically precise
results has proven valuable. For example, the most accurate calculations of the Higgs-boson cross section includes contributions from logarithmically
enhanced terms derived using a factorization formula~\cite{deFlorian:2009hc,Ahrens:2010rs}. Other examples are the inclusive photon~\cite{Kidonakis:1997gm,Laenen:1998qw,Becher:2009th,Becher:2012xr}
 or $W$-boson~\cite{Becher:2011fc,Kidonakis:2012sy} transverse momentum spectrum, the $t\bar{t}$ cross section~\cite{Cacciari:2011hy,Ahrens:2011px}, and jet shapes such as jet mass~\cite{Chien:2012ur,Dasgupta:2012hg,Jouttenus:2013hs} or $n$-subjettiness~\cite{Thaler:2010tr,Feige:2012vc}.

Even at fixed order in perturbation theory, factorization is useful.
When computing cross sections beyond leading order in perturbation theory, infrared divergences must cancel between real emission and loop diagrams. Imposing a simple infrared cutoff is not useful for numerical evaluation, since
it requires the cancellation of large positive and negative contributions. It is more efficient to evaluate
these cross sections using a subtraction scheme based on the universal behavior in the infrared singular limits, as described by a factorization formula. For example, DGPAP splitting functions describe analytically the behavior of cross sections
in collinear limits. Spin-dependent expressions for tree-level collinear, soft and soft-collinear singular regions
of cross sections can be found in~\cite{Catani:1999ss}.
 
In addition to being phenomenologically important, factorization formulae can elucidate profound structures hidden in quantum field theories. Factorization is related to the universality of the infrared structure of gauge theories. While
this universality has been explored for decades~\cite{Ellis:1978sf,Ellis:1978ty,Amati:1978wx,Amati:1978by,Bern:1994zx,Bern:1999ry,Catani:1999ss,Nagy:2003qn},
it is still not completely understood and an active area of contemporary research~\cite{Dixon:2008gr,Becher:2009cu,Gardi:2009qi,Ermolaev:2011aa}. 
With the recent resurgence of interest in on-shell approaches
to scattering amplitudes\cite{Britto:2005fq,ArkaniHamed:2012nw},
it is natural to ask whether an on-shell approach can shed light new light on factorization.
In this paper, we give a preliminary affirmative answer to this question. 
Although we only work at tree-level, considering on-shell final-state particles using spinor helicity methods, we will see that the on-shell
approach clarifies some aspects of factorization which are buried in the formalism of other approaches.

The Collins-Soper-Sterman approach to factorization (which is entirely perturbative)
begins by  identifying regions of real or virtual phase space which can produce singularities in Feynman
diagrams~\cite{Libby:1978qf,Sterman:1978bi,Collins:1989gx,Collins:1988ig}.
These singularities, sometimes called pinch singular surfaces, are the solutions to the Landau equations \cite{Landau:1959fi}. They come from
vanishing denominators of
Feynman amplitudes, dependent on the topology of the graph, but largely independent of the
theory. 
Since these singularities are due to long-distance physics,
compartmentalizing them into sectors which are separately finite implies that there is no long-distance
communication among sectors, the hallmark of factorization. 
In this approach, jets
are abstract objects 
identified with a small region of size $\lambda$ around nonzero-momentum solutions to the Landau equations.
With factorization proven, hard, jet,  and soft functions, as well as nonperturbative objects, such as  parton-distribution and fragmentation  functions, can be defined precisely. However, it appears not to be critical to connect the operator definitions for these objects to the factorization proof itself.

An alternative approach to factorization is provided by Soft-Collinear Effective Theory (SCET)~\cite{Bauer:2000ew,Bauer:2000yr,Bauer:2001yt,Beneke:2002ph}.
In SCET, at least as applied to collider physics, the procedure has so far been more practical. 
In SCET one assumes, often without a completely rigorous proof, that factorization holds
and then derives formulae for cross sections in terms of gauge-invariant
matrix elements of effective-theory fields with interactions different from those of
full QCD.
The Lagrangian for SCET is derived by power-counting at the level of operators, rather than diagrams (as with the Landau-equations).
Then one uses similar power counting to derive factorization formulae, with the appropriate operators coming out automatically.

Unfortunately, some of the steps in the derivation of SCET are unintuitive and perhaps-unnecessary.
 For example, suppose we are interested in a process with a jet going in the $n^\mu$ direction. Then an energetic gluon in this jet should have a momentum
$p^\mu$ which is collinear to $n^\mu$, so, $p^\mu \sim E n^\mu$ for some $E$. 
 However, Lagrangians are expressed in terms of fields, not momenta. Thus, 
the {\it label-formalism} approach to SCET~\cite{Bauer:2000ew,Bauer:2000yr,Bauer:2001yt} begins by assigning scaling behavior  to quark 
and gluon fields.\footnote
{
The alternative {\it multipole expansion} approach~\cite{Beneke:2002ph,Beneke:2002ni}, position space rather than momentum space takes a more prominent role.
}
 In~\cite{Bauer:2000yr}, it is argued that a gauge field $A^\mu(x)$ associated with a collinear direction
 should scale like a collinear momentum $A^\mu(x) \sim p^\mu$.
Indeed,
 if the effective theory is to be gauge invariant at some order in a power counting parameter, both terms in $D^\mu = \partial^\mu - igA^\mu $ should have the same power counting, so this is a natural choice.
  On the other hand, the gauge field acts on a gluon state 
 $|p,h \rangle$  with momentum
 $p ^\mu$ and helicity $h$  as
 \begin{equation}
 \bra{0} A^\mu(x) \ket{p,h} = e^{-ip\cdot x}\,\epsilon_h^\mu(p) 
\end{equation}
with  $\epsilon_h^\mu$ the polarization vector.
Thus, assigning the scaling behavior $A^\mu \sim p^\mu$ forces the polarization vector
to scale like a collinear momentum $\epsilon^\mu \sim p^\mu$. This is the
scaling of an unphysical, longitudinal polarization! 
Moreover, the gauge transformations $A_\mu \to A_\mu + \partial_\mu \alpha + \cdots$
which are consistent with the scaling $A^\mu \sim p^\mu$ are the limited set for which $\partial_\mu \alpha(x) \sim p^\mu$.
Thus, the derivation of SCET in~\cite{Bauer:2000yr} takes place in a particular subset of gauges.
As we will see in Section~\ref{sec:sQED}, this set of gauges, where polarizations are nearly longitudinal, is 
the only one where an intuitive, semi-classical picture of gauge-boson emission does {\it not} apply.
Of course, there is nothing wrong with 
considering a reduced class of gauge transformations in deriving an effective theory. 
However, such a strange gauge limits our ability to apply semi-classical intuition to see why
factorization holds.

In contrast, traditional approaches to QCD, such as~\cite{Catani:1999ss}, tend to discuss factorization in ``physical gauges'', where the polarization vectors are not nearly longitudinal. In physical gauges, semi-classical intuition does
apply. However, why factorization should hold in an unphysical gauge, such as the ones used in SCET, is not obvious
in the traditional approach. Also, these traditional approaches
tend to describe universality at the cross-section, rather than
amplitude level. In summing over outgoing polarizations, the fact that factorization holds at the amplitude level
is obscured.


The goal of this paper is to give a more transparent derivation of factorized forms for
matrix elements in perturbative QCD.
We focus on finite real-emission graphs with physical on-shell states (rather than on the singularities of loops which are connected only indirectly to real-emission
graphs through unitarity). Although
gauge-invariance will play a critical role, we will not have to choose a particular gauge, and we will not have to assign
a scaling behavior to unphysical fields.

The main result of this paper is a rigorous and self-contained tree-level derivation of a factorization formula in perturbative QCD in which all the objects involved as well as the expansion parameter $\lambda$ have transparent definitions from the beginning. What we will show is that
 \begin{equation}
\bra{X_1\cdots X_N; X_s} \bar\psi  \cdots \psi
\ket{0}
\LPeq
\bra{X_1} \bar{\psi} W_1 \ket{0}\cdots
\bra{X_N} W_N^\dag \psi \ket{0}
\bra{X_s} Y_1^\dg  \cdots Y_N\ket{0} \label{overview}
\end{equation} 
Here, $\bra{X_j}$ are states involving collinear momenta, all traveling in cones of opening angle $\lambda$,
and $\bra{X_s}$ are states involving soft momenta satisfying $k^\mu \lesssim  \lambda^2 Q$, where $Q$ is the center-of-mass energy of the entire state $\bra{X_1\cdots X_N; X_s}$.
The $W_j$ and $Y_j$ are Wilson lines which we define in Section~\ref{sec:Wilson}. All fields are evaluated at a common point $x=0$.
The $\LPeq$ symbol indicates that the two sides agree at leading non-vanishing order in a series expansion in $\lambda$. Color indices are suppressed. All matrix elements are taken in full QCD.
 We will prove \Eq{overview}, and its generalizations, rigorously at tree level at leading non-vanishing order in $\lambda$.

A tree-level factorization of collinear and soft states similar to \Eq{overview} was discussed in \cite{Catani:1999ss}, writing the result instead for the squared matrix element in terms of splitting functions and soft currents.
However, a factorization formula in terms of matrix elements of gauge-invariant QCD operators in a form close to that of \Eq{overview} has only appeared in the literature rather recently~\cite{Freedman:2011kj,Freedman:2013vya}. Earlier incarnations, where fields are not QCD fields but effective field theory fields, sprinkle the SCET literature.
Indeed, once \Eq{overview} is proven, the derivation of SCET (or rather, a theory which is equivalent to SCET at leading power) is almost trivial. 
We define the effective theory Lagrangian as $N+1$ copies of the QCD Lagrangian
\begin{equation}
\LEFT = \cL_1 + \cdots \cL_N + \cL_{\text{soft}}
\label{SCET1}
\end{equation}
This decoupled Lagrangian is the one proposed in the simplified formulation of SCET in \cite{Freedman:2011kj}.  Then, assigning quantum numbers to the states in the different sectors one can collapse the various matrix elements back to a simple form
 \begin{equation}
\bra{X_1\cdots X_N; X_s} \bar\psi_1  \cdots \psi_N
\ket{0}_\LQCD
\LPeq
\bra{X_1\cdots X_N; X_s} \big(\bar{\psi}_1 W_1 Y_1^\dg\big) \cdots \big(Y_N W_N^\dg\psi_N \big) \ket{0}_\LEFT
\label{SCET2}
\end{equation} 
The advantage of doing this is that now one has an equivalence of a matrix element in QCD
and a matrix element of a single operator in an effective theory. Thus, for example, 
one can compute the anomalous dimension of this operator to resum logarithms.

Returning to \Eq{overview}, we want to emphasize that this factorization formula is a statement about amplitudes in perturbative QCD. Although effective
field theory techniques can be used to make formulae like this extremely predictive and powerful, they are not critical to understanding factorization 
or expressing it in a concise form.

To prove \Eq{overview}, which we do at tree-level,
we will not have to make any assumptions about the spins of external states, or the scaling behavior of fields. A key tool which makes this possible is the spinor-helicity formalism. With spinor-helicity methods, lightlike four-momenta can be written as
an outer product of spinors, $p^\mu = p \rangle [ p$. Polarization
vectors, $\epsilon^\mu(p)$, of definite helicity are also lightlike and can be written as an outer product of a spinor associated
with their momentum, $p]$ or $p\r$, and a spinor, $ r\rangle$ or $r ]$, associated with any other lightlike four-vector $r^\mu$ called the {\it reference vector}.
The reference vector $r^\mu$ and the momentum $p^\mu$ define
 a plane to which the polarization vector is orthogonal $p\cdot \epsilon=r\cdot\epsilon=0$.
 Since the reference vector can be chosen to be some momentum in the external state
 of interest, we can write any matrix element in QCD in terms of helicity spinors associated
 with physical on-shell momentum. This lets us use only the scaling of external momenta
 to simplify matrix elements at leading power.

 An additional advantage of spinor-helicity methods is that one can choose reference vectors differently
 for different external states. 
 Matrix elements are invariant under choice of reference vector $r^\mu$ if and only
 if they satisfy the Ward identity.
The choice of reference vectors $r^\mu$ is in a way similar to a choice of gauge. For example, different $r^\mu$ can move dominant contributions from one Feynman diagram to another, and only the sum is $r^\mu$-independent (as only the sum of diagrams is gauge invariant). However, gauge choices are associated with fields. For example, trying to have $n_j\cdot A=0$ for fields associated with
gluons collinear to different $n_j$ directions, becomes a highly 
nonlocal\footnote{If one changes to radial coordinates, this particular gauge choice becomes
local; it is just Coulomb gauge in AdS~\cite{Chien:2011wz}.} 
constraint on  $A_\mu$.
In contrast, there is nothing awkward about choosing $r^\mu$ differently
in different collinear sectors. 
Indeed, this freedom of choosing $r^\mu$ dramatically simplifies the calculation of even the simplest gluon scattering amplitudes. It will also simplify our proof of \Eq{overview}.

The on-shell approach to factorization clarifies the role played by the different ingredients, such as the soft and collinear Wilson lines, in \Eq{overview}. We therefore have organized this paper by considering factorization in  field theories of increasing complexity. Section~\ref{sec:orient} provides an orientation to our approach. It includes a motivation of why soft and collinear states are relevant, from the point of view of on-shell states, and reviews spinor-helicity methods and power-counting. We also provide a review of the Wilson lines in this orientation. 
In Section~\ref{sec:scalar} we explain why scalar field theories do not factorize. The simplest theory that does admit a factorization formula like \Eq{overview} is scalar QED, which is the subject of Section~\ref{sec:sQED}. Scalar QED already contains much of the relevant physics that goes into the soft and collinear limits, so we pause to discuss the physical picture of soft-collinear factorization in Section~\ref{sec:physpict}.
To the extent that factorization can be understood from the tree-level considerations in this paper, it can be understood in scalar QED.
With the general result proven for matrix elements of a specific operator, we show how it applies to any hard process, including those with identical particles, in Section~\ref{sec:IdentPart}.

Going from scalars to spinors in Section~\ref{sec:QED} is straightforward and elucidates some new elements of factorization, such as spin-independence of the soft limit. The generalization to QCD is given in Section~\ref{sec:QCD}, where
gluon self-interactions further illustrate some aspects of factorization. 
We briefly compare our results to SCET in Section~\ref{sec:SCET}, which contains little more than a repetition of Eqs.~\eqref{SCET1} and \eqref{SCET2}.
As an application of the on-shell approach to factorization, we give in Section~\ref{sec:splitting} a concise derivation of the Altarelli-Parisi splitting functions
in QCD. In particular, due to the factorization formula, we not only derive the splittings functions but show that they apply to any process. 
Section~\ref{sec:conclusions} discusses some conclusions and provides a brief outlook.


\section{Orientation \label{sec:orient}}

Our first task is to establish precisely what we mean by factorization, and in what limit we expect it to hold. This section establishes the importance of soft and collinear limits and the notation of lightcone coordinates.  It then reviews some aspects of the spinor-helicity formalism, and shows how it can be used to power-count expressions involving polarization vectors.

\subsection{Power counting momenta}
The type of factorization we discuss in this paper applies to QCD in processes with clearly separated jets of collinear radiation. These jets can be incoming; for example, a proton can be thought of as an incoming collection of collinear radiation. But for simplicity, we focus on processes with outgoing jets only,
such as $e^+e^- \to \text{jets}$. Actually, since we will only discuss quarks and gluons, let us write the
process as $e^+e^- \to X$, with $\langle X|$ a generic partonic final state.

One can always partition any final state $\langle X |$ into collections $\langle X_i |$, for
example with a jet algorithm. For each partition one can sum all the quark and gluon
energies into the jet energy $Q_i$ and compute the jet mass $m_i$ from the sum of the
parton four-momenta. Then a power counting parameter can be defined as
$\lambda_i\equiv m_i/Q_i$. A partition with $\lambda_i=0$ is  massless, thus it can consist of only a single parton, or a set of partons which are exactly collinear to each other. Small but nonzero $\lambda_i$ mean that the partitions look like collimated collections of particles which are conventionally called jets (we will not need a precise jet definition to show factorization).

What we want to show is that the matrix element for producing any state $\langle X |$ can be computed using a factorized expression when all of the $\lambda_i$ are small.
To get an idea of what this means, consider what kind of final-state partition can produce a small but non-zero $\lambda$.
For $\lambda$ to be non-zero, a partition must have at least two partons.
Say it has exactly two partons with momenta $p^\mu$ and $k^\mu$ then the jet mass is
\begin{equation}
m^2 =(p+k)^2 = 2 p \cdot k = 4 E_p E_k \sin^2\frac{\theta}{2} \label{m2def}
\end{equation}
where $\theta$ is the angle between the 3-vectors $\vec p$ and $\vec k$.
 Thus, $\lambda = \frac{m}{E_p+E_k} \ll 1$ when either one of energies is small or when the angle between the two is small. More generally, if there are many particles in the jet, then
 $\lambda \ll 1$ if and only if the jet contains only particles which are all either soft or close in angle to the same direction (collinear).\footnote
{
These soft and collinear regions are of course the same ones which characterize solutions to the Landau equations when external momenta are massless.
On the other hand, characterizing the regions through properties of the momenta in the jets, as we have done, avoids any discussion
of Feynman diagrams which is more consistent with the on-shell approach.
}

  To be concrete, let us define the jet momentum  as the sum of the momenta of the partons in the partition:  $p_J^\mu = \sum_{i \in \text{partition}} p^\mu_i$. Then we can define the light-like four-vectors $n^\mu = (1, \vec p_J/|\vec p_J|)$ and $\bar n^\mu = (1, -\vec p_J/|\vec p_J|)$. For a jet with a single parton of momentum $p^\mu$,
 then
  $E_p n^\mu = p^\mu$ exactly. We can write any four-vector $V^\mu$ in light-cone coordinates as
  \begin{equation}
V^\mu = \frac{1}{2} V^+ \bar{n}^\mu + \frac{1}{2}  V^- n^\mu + V_\perp^\mu
  \end{equation}
where $V^+ =n\cdot V$ and $V^- = \bar{n} \cdot V$. 
Writing $p^\mu = (p^+,p^-, p^\mu_\perp)$ and
 $k^\mu = (k^+,k^-, k^\mu_\perp)$,
 and setting the jet energy $Q=E_p+E_k=\frac{1}{2}(p^++p^- + k^+ + k^-)=1$ for simplicity,
\Eq{m2def} becomes
\begin{equation}
m^2 =(p+k)^2 = p^+ k^- + p^- k^+ + 2p_\perp\cdot k_\perp  = \lambda^2 
\label{m2lam}
\end{equation}
It follows that for $\lambda \ll 1$, at least one of the momenta must have {\bf collinear scaling}:
\begin{equation}
 p^\mu = (p^+,p^-, p^\mu_\perp) \sim (\lambda^2,1,\lambda) 
\end{equation}
and the other can either have collinear scaling or {\bf soft scaling}:
\begin{equation}
 k^\mu = (k^+,k^-, k^\mu_\perp) \sim (\lambda^2,\lambda^2,\lambda^2)
\end{equation}
These $\sim$ relations mean that components can be smaller than the expressed power of $\lambda$, but not larger (up to perhaps a factor of order one).
It may seem that \Eq{m2lam} would be consistent with both $k^\mu$ and $q^\mu$ scaling like $(\lambda,\lambda,\lambda)$, however such scaling would be inconsistent with our normalization $Q=1$.\footnote
{In the literature $(\lambda,\lambda,\lambda)$ is often called soft scaling, and $(\lambda^2,\lambda^2,\lambda^2)$ is called ultrasoft. We will only be concerned with the latter and will call it simply soft. 
} 
We will use the following notation throughout this paper: $q\parallel p$ means that $q$ is collinear to $p$, which implies  $q\cdot p \sim \lambda^2$, and we will drop the $Q$ when writing the scaling of momenta in terms of $\lambda$.

For any number of momenta in a jet, if any two have scaling which is not collinear to that jet direction or soft, the jet mass will be larger than $\lambda$.
 We therefore conclude that if we split the momenta in a state $\langle X |$ into $N+1$ partitions, all of which have $\lambda_i \ll 1$
and no two have $\lambda \ll 1$ when combined, then $N$ of the partitions must be $\langle X_i|$ which are collinear to some direction $n_i^\mu$ and the 
remaining one must be soft $\langle X_s|$. That is,
\begin{equation}
\langle X | = \langle X_1  X_2 \cdots X_N ; X_s |
\end{equation}
We will refer to states of this form as $N$-jet states. This decomposition does itself not imply factorization; it is just a statement about phase space.
The leading order state 
has one parton in each state, 
$\bra{p_1 p_2 \cdots p_N}$, where $p_i$ are the momenta of the partons in the various jets and there is no soft momentum.
States also have helicities, but we suppress helicity indices for simplicity.
The general state $\langle X |$ with $\lambda_i \ll 1$ can have additional collinear momenta, which we denote $q_{a_1} \cdots q_{b_1}$ in the first jet, $q_{a_2} \cdots q_{b_2}$ in the second jet, and so on. It can also have particles with soft momenta $k_1\cdots k_\ell$:
\begin{equation}
\langle X | = \langle p_1 \cdots p_N ; q_{a_1} \cdots q_{b_N}; k_1 \cdots k_\ell |
\end{equation}
The generalization to include some incoming and some outgoing particles only amounts to keeping track of appropriate signs.

What we will show is that the matrix element  $\langle X | \cO |0 \rangle$ of a \emph{hard-scattering operator} $\cO$ can be written in a factorized form:
\begin{equation}
\langle X | \cO |0\rangle 
\LPeq \langle p_1;q_{a_1} \cdots q_{b_1}| \cO_1 |0\rangle
\cdots
\langle p_N;q_{a_N} \cdots q_{b_N}| \cO_N |0\rangle
\langle k_1 \cdots k_\ell | \cO_s |0\rangle
\label{heuristicO}
\end{equation}
at leading order in $\lambda_i $
 for some suitable collinear operators $\cO_i$ and soft operator $\cO_s$.
We assume for simplicity that the hard-scattering operator $\cO$ has $N$ fields, one for each collinear direction,
with no two fields describing identical particles, and that $\cO$ has
a non-vanishing matrix element in the leading order state $\bra{p_1 p_2 \cdots p_N}$. 
For example, $e^+e^- \to$ dijets is mediated by a combination
vector and axial current operators $\cO_V = \bar{q} \gamma^\mu q$ and $\cO_A = \bar{q} \gamma^\mu \gamma_5 q$.
Other examples can be found in~\cite{Kelley:2010fn} or~\cite{Stewart:2012yh}.

It is not actually necessary for a matrix element to be expressible in terms of 
a hard-scattering operator for factorization to hold. As long as the process 
involves $N$ different directions, the amplitude can factorized at leading power into a sum of
expressions of the form on the right-hand-side of \Eq{heuristicO}. We choose to express factorization in the
form of  \Eq{heuristicO} in the bulk of this paper, with $\cO$ having $N$ fields each with a unique flavor
quantum number, mostly to have cleaner looking equations and to avoid nettlesome combinatoric issues. 
A demonstration that the factorization we derive
applies to more general hard scattering processes, such as those with identical particles, is given in Section~\ref{sec:IdentPart}.
The only strict requirement for factorization is that no two sectors are collinear. In particular, in cases like forward
scattering or Drell-Yan, where
incoming partons may be collinear to outgoing spectators, more care is needed to prove factorization\cite{Catani:2011st}. We do not address such
cases here.

Regardless of the connection between operator matrix elements and scattering amplitudes,
\Eq{heuristicO}, when made more precise, is a highly nontrivial statement about
how Yang-Mills theories simplify. We will therefore eschew any additional discussion
 of the connection to infrared safe observables and $S$-matrix elements,
focusing instead on perturbative factorization 
of the given matrix elements.


\subsection{Spinor helicities and power counting}
\label{sec:spinorhelicity}

Matrix elements in theories with spin-1 particles produce expressions involving polarization vectors. For example, they
may contain terms like $p \cdot \epsilon$ or $\epsilon_1 \cdot \epsilon_2$. In order to know which of these terms are dominant at small $\lambda$, we need to power-count polarizations. 
These products might vanish for some
polarizations, changing which diagrams contribute at the leading power. 
There are in fact many ways to represent the same physical
photon helicity state with polarization vectors, so that the scaling $p \cdot \epsilon$ 
is not even well-defined given particular states in the Hilbert space.
This scaling behavior of expressions involving polarization vectors
is easiest to derive if we represent the polarizations in terms of helicity spinors.

In the spinor-helicity formalism, massless left and right handed spinors (i.e. solutions to $\Sl{p}\,u(p)=0$) are written as
\be
u_L(p)^\a = p\r \quad\And\quad u_R(p)_\ad = p]
\ee
where the $\a$ and $\ad$ indices are those of two separate $SU(2)$'s. This way, $SU(2)$-invariant spinor products can be written as
\be
\l pk\r = \epsilon^{\a\b} \,\l p_\a\, \l k_\b = \l p_\a\, k\r^\a \quad\And\quad
[ pk ] = \epsilon^{\bd\ad} \,p]_\ad\, k]_\bd = [ p^\bd\, k]_\bd
\ee
Four-vectors live in the $(\frac12,\frac12)$ representation of the Lorentz group, so they may be expressed with these indices as well, via
\be
p^{\a\ad} = \sigma_\mu^{\a\ad} \,p^\mu \quad\And\quad
p^\mu = \frac12 \bar\sigma^\mu_{\ad\a}\,p^{\a\ad}
\ee
In this representation, masslessness of a four-vector becomes 
\be
p^2 = \det p^{\a\ad} = 0
\ee
which implies
\be
p_{\ad\a} = p]\l p  \quad\And\quad
p^{\a\ad} = p\r[p 
\ee
and
\be
2p\cdot k = p_{\ad\a}k^{\a\ad} = [ kp]\l pk\r
\ee

Polarization vectors of a definite helicity are also massless. They can be defined by their normalization: $\epsilon_q^2 = 0$ and $\epsilon_q^*\cdot\epsilon_q = -1$, transversality: $q\cdot\epsilon_q = 0$ and one more condition: $r\cdot \epsilon_q = 0$ for $r^\mu$ linearly independent of $q^\mu$. 
The four-vector $r^\mu$ is called the {\bf reference vector}.
The above constraints are elegantly encoded with helicity spinors as follows:
\begin{equation}
    \epsilon_q^-(r)=
    \sqrt{2}\, \frac{q \rangle [ r}{[ q r ]}
    \quad\And\quad
     \epsilon_q^+(r) = \sqrt{2}\, \frac{r \rangle [ q }{\langle r q    \rangle}
\end{equation}
Thus, for example, 
\be
\frac{p\cdot\epsilon^-_q}{p\cdot q } = \sqrt{2}\, \frac{[rp]}{[qr][qp]}  
\ee
The freedom of choice of reference vector encodes the arbitrariness associated with assigning a specific polarization vector to a given helicity. Consequently, matrix elements that satisfy the Ward identity will be independent of the choice of reference vector.

To see how spinors and spinor products scale with $\lambda$, first recall that there is not a unique way to write a massless four-vector in terms of spinors: any little group transformation $p] \to p] z$ and $\langle p  \to \frac{1}{z}\langle p$ for any complex number $z$ leaves $p]\langle p$ unchanged. For real momenta, $|z|=1$. Since $k^\mu \sim \lambda^2$ for a soft momentum, the associated spinors must scale like $k] \sim \langle k \sim \lambda$. Moreover, soft momenta do not have a specific direction, so, without loss of generality, we can choose the reference vectors for soft-photon polarizations to satisfy $[kr],\l rk\r \sim 1$. The scaling of the soft polarization vectors is then fixed to be $\epsilon_k^\mu \sim 1$.

For collinear momenta, the components of the helicity spinors do not have uniform scaling; nor, therefore, do the polarization vectors. However, since for two collinear four-vectors $2 p\cdot q =  \l qp\r[p q]\sim\lambda^2$ and since $\l qp \r= [p q]^*$, we conclude that $\l qp \r \sim [p q] \sim  \lambda$. Thus we will be able to power count Lorentz-contracted products involving collinear momenta and polarizations.


\subsection{Wilson lines\label{sec:Wilson}}
Wilson lines play an important role in factorization. They describe the radiation produced by a charged particle
moving along a given path in the semi-classical limit. The semi-classical limit applies when 
the back-reaction of the radiation on the particle can be
neglected. In particular, this limit holds when the particle is much more energetic than any of the photons in the radiation, that is,
when the photons are all soft. Thus Wilson lines naturally appear in the soft limit of Yang-Mills theory. That they also play
a role in collinear limits is less obvious and will be explained in Section~\ref{sec:sQED}.

An outgoing Wilson line in the $n^\mu$ direction is defined by
\begin{equation}
  Y_{n}^\dag(x) = P \left\{ \exp \left[ i g 
  \int_0^{\infty} d s \,n\cdot A ( x^{\nu} + s n^{\nu} )\, e^{-
  \varepsilon s} \right] \right\} \label{softYQCD}
\end{equation}
where $P$ denotes path-ordering and $A = A^a T^a$ is the gauge field in the fundamental representation (Wilson lines in other representations are a straightforward generalization).
This Wilson line is outgoing because the position where the gauge field $A_\mu(x)$ is evaluated goes from
$x$ to $\infty$ along the $n^\mu$ direction. We write $Y_n^\dag$ for Wilson lines for outgoing particles,
and $Y_n$ for outgoing antiparticles (as $\bar{\psi}$ creates outgoing quarks and $\psi$ creates outgoing antiquarks). 
Explicitly,
\begin{equation}
  Y_{n}(x) = \overline{P} \left\{ \exp \left[- i g
  \int_0^{\infty} d s\, n\cdot A ( x^{\nu} + s n^{\nu} ) \, e^{-
  \varepsilon s} \right] \right\}
\end{equation}
where $\bar{P}$ denotes anti-path ordering.
We write incoming Wilson lines as
\begin{align}
  \bar{Y}_n(x) &= P \left\{ \exp \left[ i g
  \int_{- \infty}^0 d s\, n\cdot A ( x^{\nu} + s n^{\nu} ) \,
  e^{\varepsilon s} \right] \right\} 
\\
  \bar{Y}^\dag_n(x) &=\overline{P} \left\{ \exp \left[ -i g
  \int_{- \infty}^0 d s\, n\cdot A ( x^{\nu} + s n^{\nu} ) \,
  e^{\varepsilon s} \right] \right\} 
\end{align}
where now the path goes from $-\infty$ to $x$ and the $ i\varepsilon$ prescription is switched.

One can have Wilson lines in any representation. For example, an adjoint Wilson line can be written as
\be
\cY_n^\dg(x) = P\bigg\{ \exp\bigg[ ig \int_0^\infty ds\, n\cdot A_\mu^a (x + s\, n) T^a_\text{adj} \, e^{-\epsilon s} \bigg] \bigg\}
\ee
where $(T^a_\text{adj})^{bc} = if^{bac}$ are the adjoint-representation group generators. A useful relation is that, because
\be
(T_\text{adj}^c)^{ab}T^b = [T^a,T^c] 
\, ,
\ee
one can write
\be
 Y_n^\dg A^\mu_b T^b\, Y_n = A^\mu_a \,\cY_n^{ab}\, T^b
\label{YYrel}
\ee
In this way, all the relevant Wilson lines in QCD can be expressed in terms of fundamental Wilson lines or their adjoints (which
are antifundamental Wilson lines).

Although Wilson lines are non-local, their matrix elements in given external states can be evaluated order by order in perturbation theory.
For a state with a single gluon of momentum $k^\mu$ and polarization $\epsilon^\mu$, we find
\begin{align}
\bra{\epsilon(k)} Y_n^\dg(0) \ket{0} &= igT^a \int_0^\infty ds\, \bra{\epsilon(k)} n \cdot A^a(s\, n^\mu)\ket{0}\, e^{-\varepsilon s} \\
&= ig T^a\,n \cdot\epsilon^*_k \int_0^\infty ds\, e^{is(n\cdot k + i\varepsilon)}\\
&= -gT^a \frac{n\cdot\epsilon^*_k}{n \cdot k + i\varepsilon} \label{eikvert}
\end{align}
This is the form of an eikonal vertex, coming from the soft limit of a QCD interaction
\be
\fd{3.5cm}{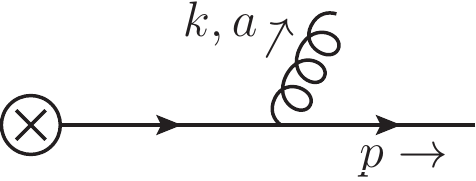} \;\;= \;-gT^a \frac{n\cdot \epsilon_k}{n\cdot k + i\varepsilon}
\ee
with the correct $i \epsilon$ prescription (and we have dropped the factor of the amplitude with no emission).
For the incoming Wilson line, a similar calculation gives
\be
\bra{\epsilon(k)} \bar Y_n \ket{0} \;=\; g T^a\frac{n\cdot \epsilon_k}{n\cdot k - i\varepsilon} 
\;=\;\;
\fd{3.5cm}{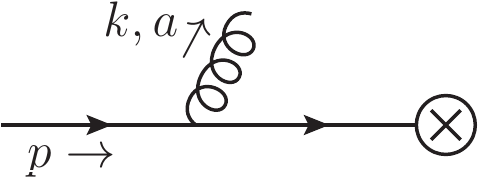} 
\ee
which also has the correct sign and $i\varepsilon$ prescription.

The $e^{\pm \varepsilon s}$ factors in these Wilson lines are required when the Wilson lines are used in time-ordered products to calculate $S$-matrix elements. This is most clearly seen in perturbation theory, where, as we have shown these $e^{\pm \varepsilon s}$ factors generate the pole displacements for Feynman propagators. 
The $e^{\pm \varepsilon s}$ factors affect the gauge transformation properties, but only at order $\varepsilon$:
\be
Y_n(x) \to e^{i \alpha(x)} Y_n(x) e^{-i \alpha(\infty)} + \cO(\varepsilon)
\label{Yinf}
\ee
Thus, for gauge transformations that vanish at infinity\footnote
{Requiring $\alpha(\infty)=0$ is not a strong restriction on the class of gauge transformations allowed, since one can always supplement a gauge transformation
with a global transformation (with $\alpha$ constant), to set $\alpha(\infty)=0$. In any case, the factorization formulas we derive
are an exact equivalence of matrix elements, at leading power. Their validity does not depend on the concept of gauge invariance.
}, $Y_n(x)$ transforms like a fundamental in the
 $\varepsilon\to 0$ limit.

It is perhaps worth noting that 
similar $\cO(\varepsilon)$ corrections are present even in a local quantum field theory. For example, in scalar QCD,
the matrix element of the operator $\phi^* \phi$ in a state $\bra{\epsilon(q) p_1 p_2} $ 
containing an outgoing gluon, scalar particle and scalar antiparticle with momenta $q$, $p_1$ and $p_2$ respectively,
is
\be
\bra{\epsilon(q) p_1 p_2} \phi^*(0) \phi(0) |0\rangle
= -g T^a\left( \frac{p_1 \cdot \epsilon}{p_1\cdot q + i\varepsilon} -  \frac{p_2 \cdot \epsilon}{p_2\cdot q + i\varepsilon} \right)
\ee
This in fact does not satisfy the Ward identity exactly, but only up to corrections of order $\varepsilon$. Indeed, substituting 
$\epsilon \to q$, we find
\be
\bra{q(q) p_1 p_2} \phi^*(0) \phi(0) |0\rangle
\overset{\epsilon \to q}= 
i \varepsilon g T^a\left(\frac{1}{p_1 \cdot q} - \frac{1}{p_2 \cdot q}\right) + \cO(\varepsilon^2)
\ee
which does not vanish exactly. Thus, the danger in violating gauge-invariance at order $\varepsilon$ is no worse when using
Wilson lines than in a theory with only local operators. In any case, since this paper is entirely about tree-level matrix elements, 
the $i \varepsilon$ prescription is irrelevant. Thus, we set $\varepsilon=0$ from now on.


\section{Scalar field theory}
\label{sec:scalar}
We begin our exploration of factorization with a simple scalar field theory with Lagrangian
%
\be
\cL = \frac12 \partial_\mu\phi\,\partial^\mu\phi + \frac{g}{3!}\phi^3
\ee
Since we only work at tree-level, the dimensionality of the coupling $g$ causes no complications.

We consider the matrix elements  $\langle X | \cO(0) |0\rangle$ of 
the hard-scattering operator
\be
\cO(x) = \frac1{N!}\phi(x)^N
\ee
in an $N$-jet state $\bra{X} = \bra{X_1\cdots X_N;X_s}$.
The normalization of the operator is chosen so that its leading-order matrix element is $\bra{p_1\cdots p_N}\cO(0)\ket{0} = 1$.
One can add derivatives to the operator with little effect on the following arguments. Indeed, adding derivatives to $\cO(x)$ simply produces an overall function of the $P_{X_i}\cdot P_{X_j} \sim Q^2$, where $P_{X_i}$ is the momentum of the $\bra{X_i}$ state, that will end up sitting out front of the factorized expressions. Because the complications of adding derivatives to $\cO$ are almost entirely notational, we will always ignore derivative insertions in this paper and describe how to treat scattering much more generally in Section~\ref{sec:IdentPart}.

We start by considering only collinear emissions.
The diagram with the $j$-th line emitting a single scalar on top of the leading-order matrix element is:
\be
\fd{3.5cm}{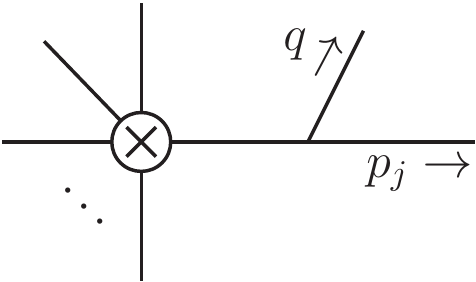}\; = \frac{-g}{2p_j\cdot q}
\label{softscalar}
\ee 
where the $\fd{.4cm}{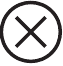}$
indicates an insertion of the operator $\cO=\frac1{N!}\phi^N$.
When $q$ is not collinear to $p_j$ then $q\cdot p_j \sim 1$, whereas,
when $q$ is collinear to $p_j$ then $q\cdot p_j \sim \lambda^2 \ll 1$.
Therefore, the diagram where the emission comes off the leg to which it is collinear 
(a {\bf self-collinear emission})
is enhanced by $\lambda^{-2}$ compared to any of the other diagrams. We can write this as
\be
\fd{3.5cm}{paperfigs/softscalar_vert.pdf} \;
\sim
\begin{cases}
\dfrac{1}{\lambda^2},&
~{q \parallel p_j} 
\\[2mm]
 1,&~{q~ \cn{\parallel} ~p_j}
\end{cases}
\label{phi3scale}
\ee 
Thus, at leading power, only the diagram with a self-collinear emission is relevant.

Since self-collinear emissions do not change the collinearity of the line from which they are emitted, the above argument can be used inductively to show that for any number of collinear particles, diagrams 
with all self-collinear emissions
are enhanced compared to other diagrams. Diagrammatically, we can write
\be
\sum\quad \fd{3.5cm}{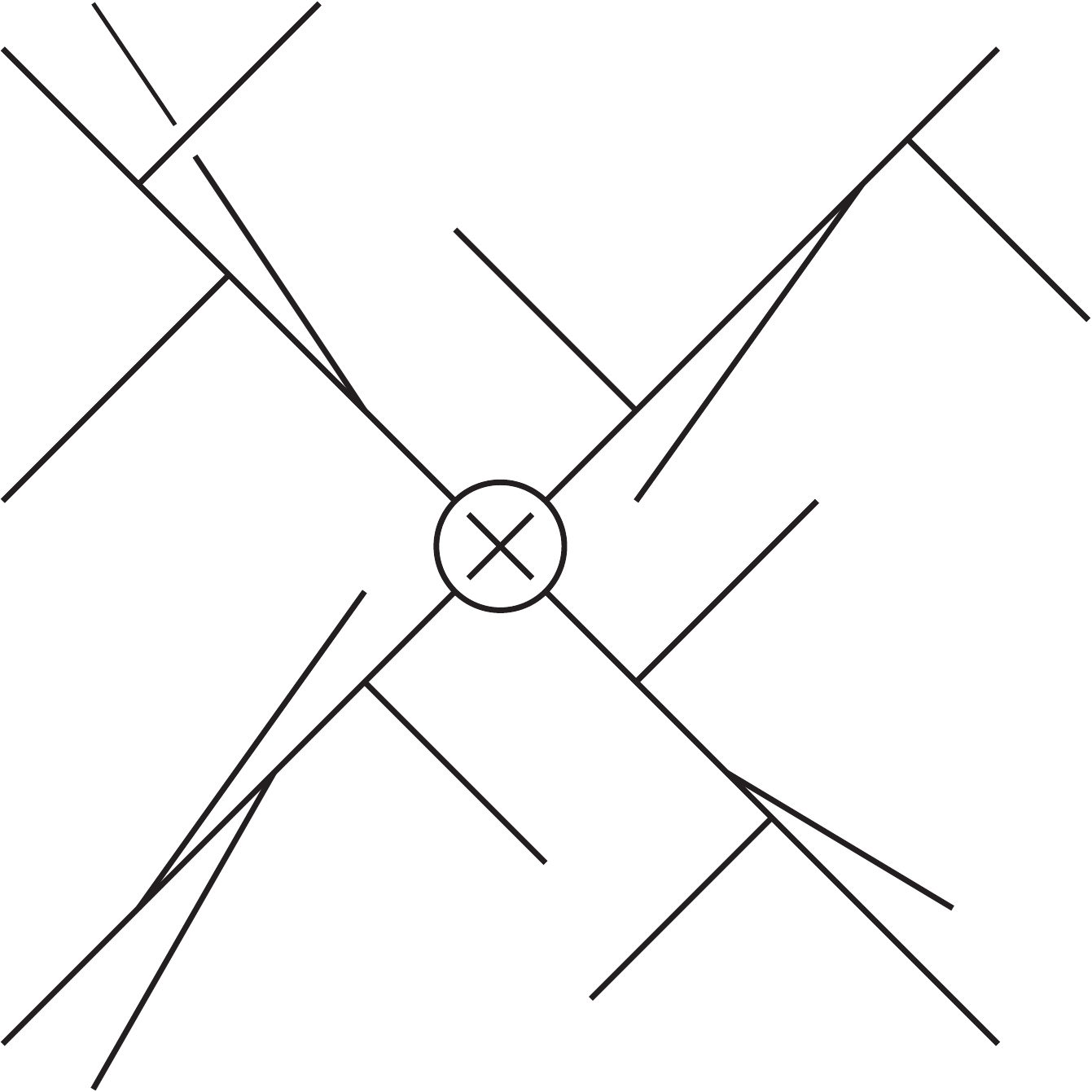}  \quad\LPeq\quad
\sum\quad \fd{3.5cm}{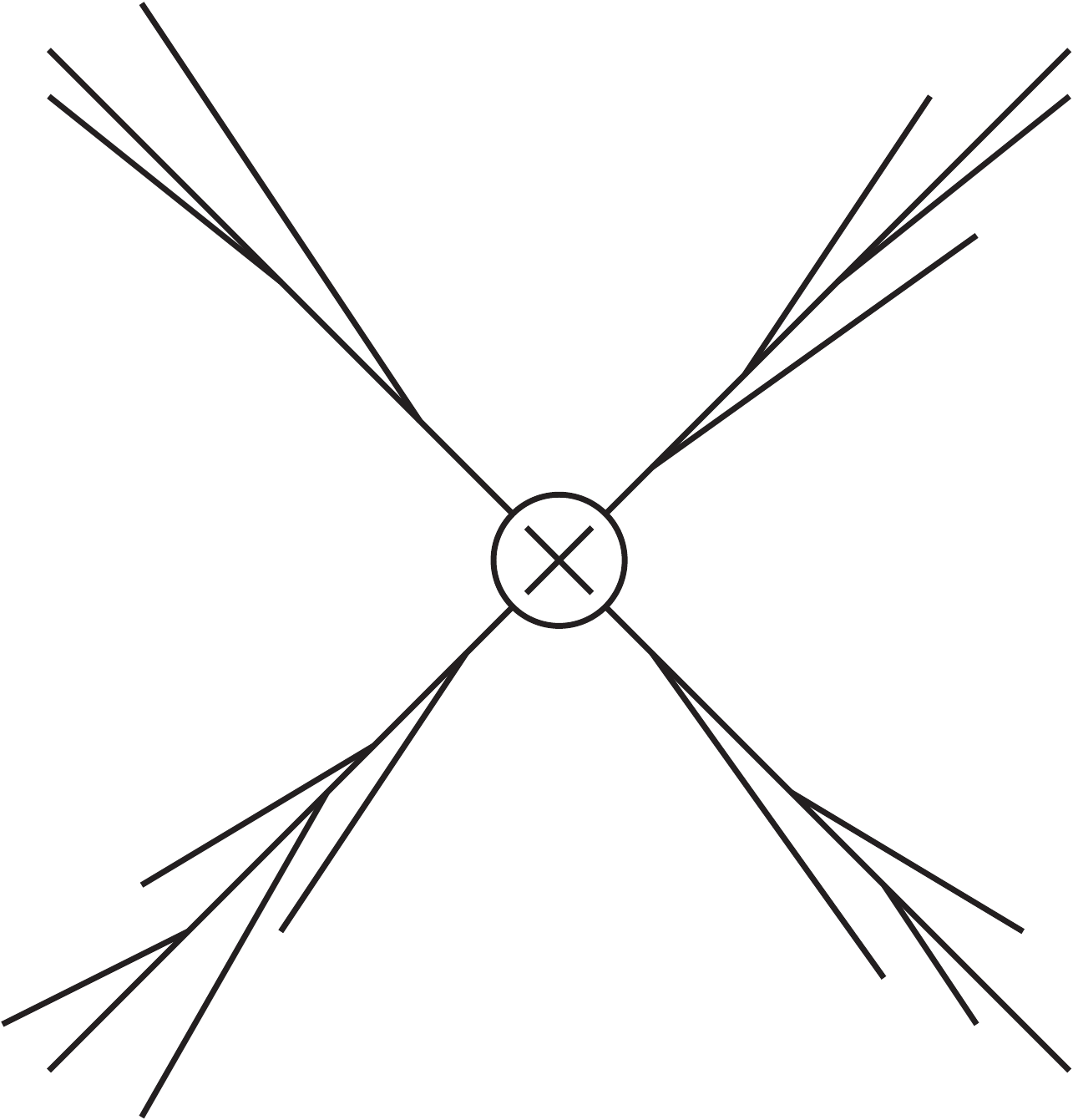} 
\label{scalarfact}
\ee
where the sum on the left means the sum of all diagrams consistent with the external state in the matrix element and the sum on the right means only those diagrams with emissions off of a line to which they are themselves collinear. The $\LPeq$ means that the two sides agree at leading power in $\lambda$.
The vertex in these diagrams denotes the hard-scattering operator of which we are taking the matrix element and for clarity, only four distinct collinear directions are shown even though we are considering $N$.

Thus, the matrix element in the scalar case simplifies when all the particles are collinear to one of the $N$ directions. It is given by the product of factors for each sector separately -- it factorizes. In terms of matrix elements, \Eq{scalarfact} can be written as
\be
\bra{p_1 \!\cdots p_N; q_{a_1}\!\cdots q_{b_N}}
\frac{1}{N!} \phi^N
  \ket{0}
\LPeq \bra{p_1;q_{a_1} \!\cdots q_{b_1}}\phi\ket{0} \cdots
\bra{p_N;q_{a_N} \!\cdots q_{b_N}}\phi \ket{0}
\ee
where $q_{a_j}\cdots q_{b_j} \parallel p_j$ for each collinear sector, $j$.
Or, more succinctly
\be
\bra{X_1 \cdots  X_N}
  \cO 
  \ket{0}
\LPeq \bra{X_1}\phi\ket{0} \cdots
\bra{X_N} \phi \ket{0}
\ee
This simplification is in fact exactly what one expects from a semi-classical picture: after the hard scattering occurs, collinear particles in each jet can be thought to emit additional scalars, as in a parton shower. There is no interference
between emissions from particles moving in different directions.

Next, consider states with soft particles.
In \Eq{softscalar}, if we take $q^\mu \sim \lambda^2$, the soft emission is also enhanced by $\lambda^{-2}$ irrespective of the direction of the soft radiation. 
Therefore, unlike the collinear case, no diagrams may be dropped. Furthermore, the Feynman rules do not particularly simplify in the soft limit, so the soft limit of the scalar theory shows no simplifications for the matrix elements under consideration.

In summary, in a scalar field theory, collinear emissions factorize and have a simple semiclassical
interpretation. However, since soft emissions do not simplify, the matrix element under consideration does not factorize
in scalar field theory at leading power.
Except in exceptional cases where soft-emission is not relevant (such as scalar $\phi^3$ deep-inelastic scattering~\cite{Collins:1989gx}), 
states with soft and collinear momenta are equally relevant to infrared-safe observables at leading power.
Thus, since all Feynman diagrams must be evaluated to reproduce the soft limit, collinear factorization by itself is not
particularly useful.


\section{Scalar QED}
\label{sec:sQED}
In scalar field theory, we saw that while self-collinear emissions dominate over collinear emissions
from distant legs, leading to an intuitive form of collinear factorization, soft emissions do not simplify in any useful way. As we will see,
in gauge theories like scalar QED, there is still  collinear factorization,
although which
diagrams dominate depends on the gauge.  There is also factorization in the soft
limit. And, most remarkably, the soft and collinear sectors factorize simultaneously.

We will be considering matrix elements of the gauge-invariant operators
 $\cO$ in the $N$-jet states discussed in Section~\ref{sec:orient}, $\bra{X} = \bra{X_1\cdots X_N;X_s}$. 
 The simplest hard-scattering operator in scalar QED, on which we focus, is
  \be
 \cO(x) = \frac{1}{(N/2)!} \left[ \phi(x)^* \phi(x) \right]^{N/2} 
 \label{simpleoperator}
 \ee 
 with $N$ even.
 Insertions of covariant derivatives in $\cO$ correspond to collinear sectors initiated by a photon; we will postpone the discussion of covariant derivatives until Section~\ref{sec:QCD} where we discuss QCD and the situation is more interesting.  
 For notational consistency with later results in QCD, we will take the QED  coupling constant to be $g = -e$.

As in the previous section, we will start our discussion using states with no soft momenta, $\bra{X} = \bra{X_1\cdots X_N}$. This will allow for a clean discussion of the essential ingredients that go into 
collinear factorization. Subsequently, we discuss states with one particle per collinear sector and many soft particles, $\bra{X} = \bra{p_1\cdots p_N;X_s}$ and finally, the simultaneous soft and collinear case, $\bra{X} = \bra{X_1\cdots X_N;X_s}$.

\subsection{Collinear Factorization}
\label{sec:sQEDCF}

Our approach to analyzing the matrix element $\bra{X_1\cdots X_N}\cO\ket{0}$ will be to start with the matrix element, $\bra{p_1\cdots p_N}\cO\ket{0}$, and to add collinear emissions until the full matrix element is constructed. We start by adding collinear photons  only and then
discuss adding additional collinear scalars. So let us take the final state of the $j$-th sector to be $\bra{X_j} = \bra{p_j,q_{a_j}\cdots q_{b_j}}$ where $p_j$ is a scalar momentum and all the $q$'s are photon momenta.

In scalar QED, the matrix element for a state with one photon is related to the matrix element for
the state with no photons as
\be \label{singemissQED}
\fd{3.5cm}{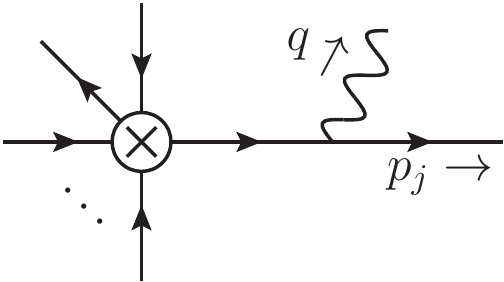} = -g\,\frac{p_j\cdot\epsilon_q}{p_j\cdot q \pie}
\ee
where the $\fd{.4cm}{paperfigs/cross_vert.pdf}$ indicates an insertion of the operator $\cO = \frac{1}{(N/2)!} | \phi |^N $ which gives a factor of $1$.
The 4-point vertex in scalar QED only contributes starting with 2 emissions: 
\be
\fd{3.5cm}{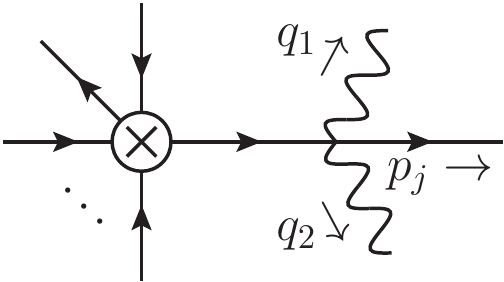} = \frac{-2g^2\,\epsilon_{q_1}\cdot\epsilon_{q_2}}{(p_j+q_1+q_2)^2\pie} 
\ee
For collinear emissions, these expressions appear to  be enhanced only when the $q$'s are collinear to $p_j$ just like in the scalar theory. However, due to the
$p \cdot \epsilon$ and $\epsilon_i \cdot \epsilon_j$ factors in their numerators, 
the story  is not so simple. 
We need to know how $p \cdot \epsilon$ and $\epsilon_i \cdot \epsilon_j$ 
scale with $\lambda$. 
For example, 
for any physical photon helicity, we can choose its associated polarization vector so that
 $p_j\cdot\epsilon_q = 0$ 
in which case the first graph would vanish (as would the second for certain helicity choices).

As discussed in Section~\ref{sec:spinorhelicity}, expressions involving
polarizations are easiest to power-count using helicity spinors.
For example, in terms of helicity spinors \Eq{singemissQED} becomes:
\be \label{singemisSH}
\fd{3.5cm}{paperfigs/collsQED_vert.pdf} = -g\,\frac{p_j\cdot\epsilon^-_q}{p_j\cdot q }
									= -g\,\sqrt{2}\, \frac{[rp_j]}{[qr][qp_j]}
\ee
where $r$ is the reference vector associated with the polarization vector, $\epsilon^-_q$ (the expression with $\epsilon^+_q$ is the complex conjugate). 
In Section~\ref{sec:spinorhelicity} we showed that the contraction of two spinors
corresponding to momenta with collinear scaling scales like the square-root
of the scaling of the contraction of the associated momenta; in equations that is $p_j\cdot q \sim \lambda^2$ implies that $[qp_j],\l p_j q\r \sim \lambda$. Thus, the scaling
of a product involving reference-vector spinors $r]$ and $[r$ can be determined
once we know in which direction $r$ points. 

 Let
us first choose reference vectors which are  
 \emph{not}  collinear to any of the momenta.
 We call this {\bf generic-$r$}. In generic-$r$,  $[rp_j]\sim [qr] \sim 1$ and
\be
\fd{3.5cm}{paperfigs/collsQED_vert.pdf}  = -g\,\sqrt{2}\, \frac{[rp_j]}{[qr][qp_j]}
\sim\frac{1}{[q p_j]} \sim 
\begin{cases}
\dfrac{1}{\lambda},&
~{q \parallel p_j} 
\\[2mm]
 1,&~{q~ \cn{\parallel} ~p_j}
\end{cases}
\label{genRoneemit}
\ee
Thus, generic-$r$ is similar to  $\phi^3$ theory (cf. \Eq{phi3scale}). However, note
 that the diagrams are less singular: one power of $\lambda$ cancels due to the scaling
 of the polarization vectors. 
In scalar QED there is also a diagram involving the four-point vertex. In generic-$r$, we find
\be
\fd{3.5cm}{paperfigs/collsQED4pt_vert.pdf} \sim
\begin{cases}
\dfrac{1}{\lambda^2},&
~{q_1\parallel q_2 \parallel p_j} 
\\[2mm]
 1,&~ 
 q_1 ~\cn{\parallel}~ p_j \Or q_2 ~\cn{\parallel}~ p_j 
\
\end{cases}
\ee
This is the same order as two emissions using the 3-point vertex (for soft emissions, as we will see, diagrams involving the 4-point vertex are power suppressed).

By induction, 
in generic-$r$, only the diagrams in which all the emissions are self-collinear
are relevant at leading power. That is, in generic-$r$,
\be
\sum\quad \fd{3.5cm}{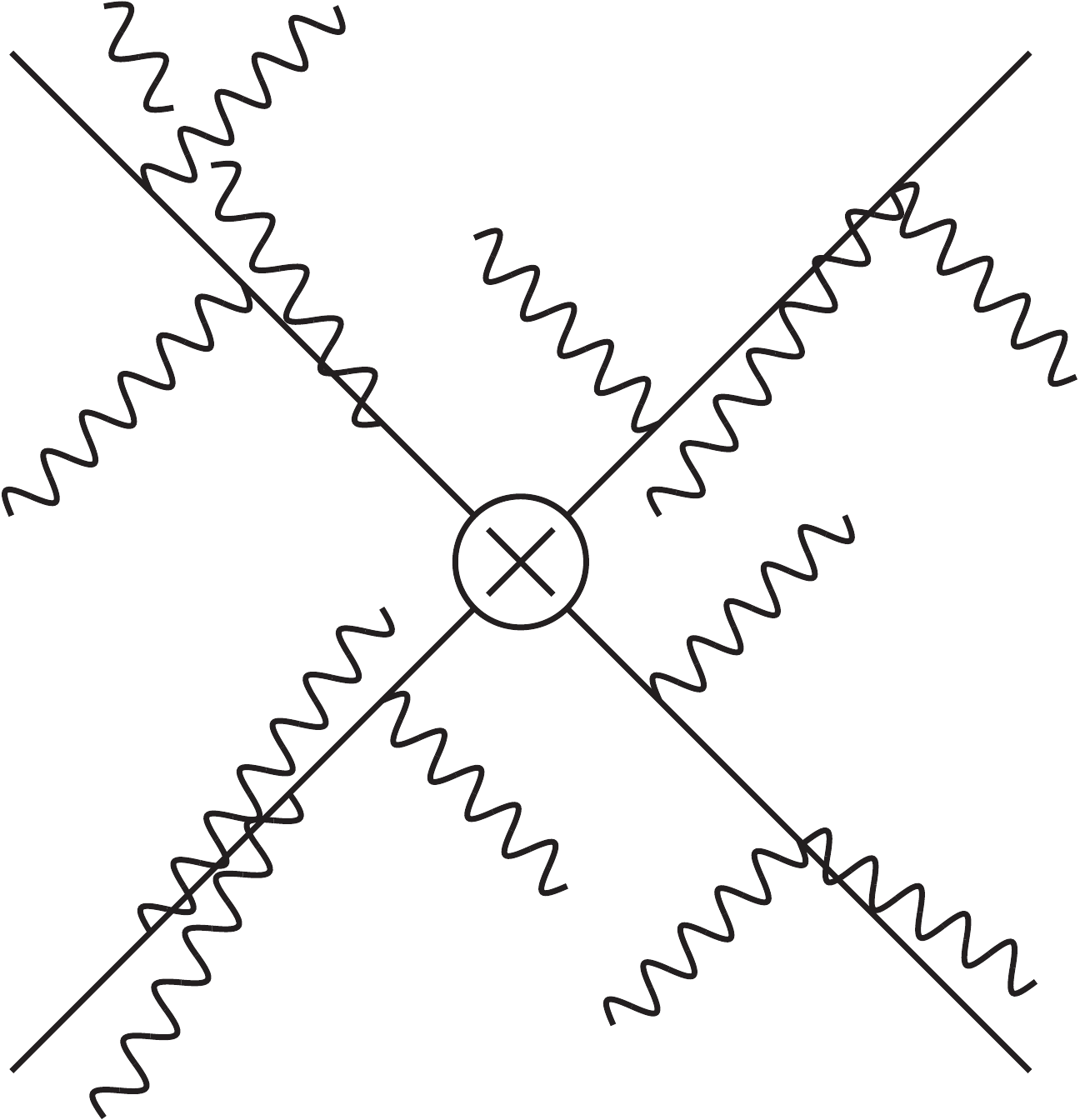} \quad\LPeq\quad 
\sum\quad \fd{3.5cm}{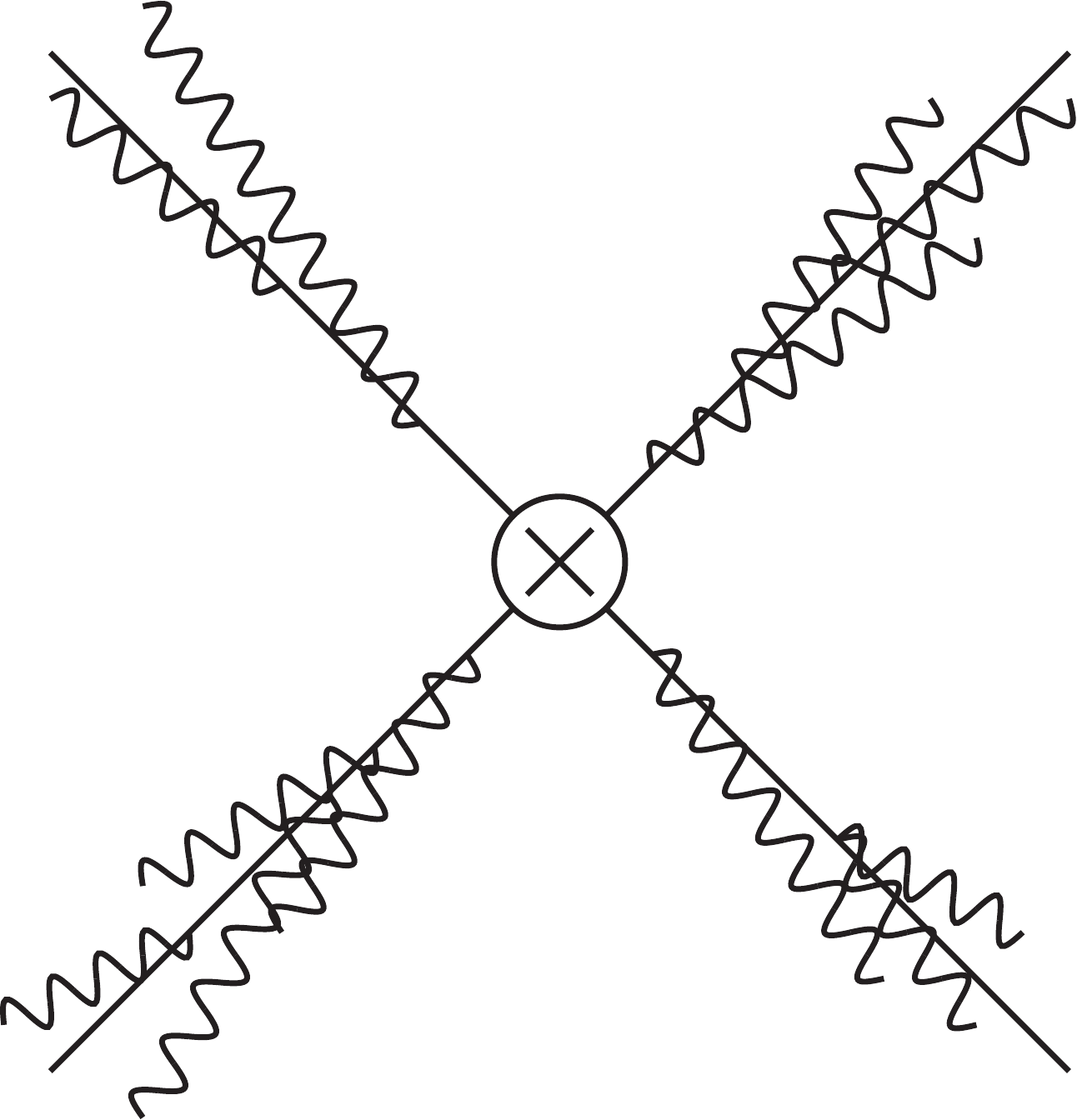} 
\label{diagfactsQED}
\ee
where the sums have the same meaning as in the scalar case: the sum on the left means the sum over all diagrams consistent with the collinear external states, namely all diagrams in $\bra{X} \cO \ket{0}$. The sum on the right means only sum over those diagrams for which the emissions in the $j$ direction are from a $j$-collinear scalar, namely, diagrams for which all emissions are self-collinear.

In terms of matrix elements, \Eq{diagfactsQED} can be written as
\be
\bra{
 p_1 \cdots p_N; q_{a_1} \cdots q_{b_N}}
\cO
\ket{0}
\LPeq 
\bra{p_1;q_{a_1} \cdots q_{b_1}}\phi^*\ket{0} \dots
\bra{p_N;q_{a_N} \cdots q_{b_N}}\phi \ket{0}
\label{matelmfactsQED}
\ee
where the collinear photons are labelled such that $q_{a_j},\cdots, q_{b_j} \parallel p_j$ for each collinear sector, $j$, and the choice of $\phi$ versus $\phi^*$ depends on whether $p_i$ is an outgoing scalar or anti-scalar. 
This equation is a precise statement of collinear factorization. It is however not gauge-invariant.
Indeed, we only showed that it holds for generic-$r$ choices of reference vector..
In fact, the sum of diagrams on the right of \Eq{diagfactsQED} does not satisfy the Ward identity
and the right-hand side of \Eq{matelmfactsQED} 
involves matrix elements of gauge-dependent fields $\phi(x)$.

As a step toward deriving a gauge-invariant form of factorization,
consider next a different choice of reference vectors.
We previously used generic-$r$ where no reference vectors could be collinear to $p_i$ for all $i$; to contrast this, we take all of the reference vectors of the $j$-collinear photons to be \emph{equal} to $p_j$, namely
\be
{\bf \text{\bf collinear-}r}:\qquad  r_{a_j},\cdots,r_{b_j} = p_j 
\ee
Then the previously-most-enhanced diagrams will be proportional to
\be
\frac{p_j\cdot\epsilon^-_q}{p_j\cdot q} = \sqrt{2}\, \frac{[rp_j]}{[qr][qp_j]} \bigg|_{r = p_j} = 0
\ee
To see where the leading-power contributions to the matrix element in \Eq{matelmfactsQED} went, note that the reference vectors for the $j$-th sector are now themselves enhanced:
\begin{equation} \label{refvectenhanced}
\epsilon_q^-(r=p_j)= \sqrt{2}\, \frac{q \rangle [ r}{[ q r ]}\bigg|_{r = p_j} = \sqrt{2}\, \frac{q \rangle [ p_j}{[ q p_j ]}
	\sim \frac{q \rangle [ p_j}{\lambda}
\end{equation}
Thus, in collinear-$r$, the leading-power contributions, those scaling like $\frac{1}{\lambda}$, all come from the non-self-collinear graphs. In diagrams, in collinear-$r$
\be
\bra{p_1\cdots p_N;q} \cO \ket{0}
\,\,=
\sum_{i\neq j}\; \fd{2cm}{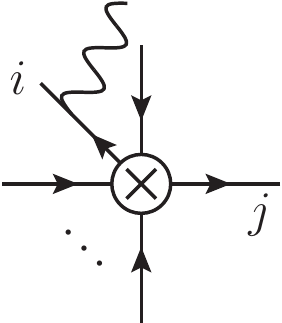} 
\ee
The sum must be the same as the self-collinear graphs which dominate in generic-$r$, since the sum of all graphs is $r$-independent.
 Interpreting this in terms of the dominant diagrams in generic-$r$ or collinear-$r$, we have found
\be
\fd{3.5cm}{paperfigs/collsQED_vert.pdf} \quad\text{in generic-}r\quad
\LPeq
\quad\sum_{i\neq j}\; \fd{2cm}{paperfigs/collsQEDi_vert.pdf} \quad\text{in collinear-}r
\label{genrcollrdiags}
\ee

It is informative to check that the sum of graphs in scalar QED is exactly $r$-independent. The sum of graphs is
\begin{align}
\bra{p_1\cdots p_N;q} \cO \ket{0}
&\,\,=\;	\fd{3.5cm}{paperfigs/collsQED_vert.pdf} \;+\;
	\sum_{i\neq j}\; \fd{2cm}{paperfigs/collsQEDi_vert.pdf} 
\\&\,\,= \sum_{i} (-Q_i g) \frac{p_i\cdot\epsilon^-_q}{p_i\cdot q} 
\\&
= -g \,\sqrt{2}\, \sum_{i} Q_i \frac{[rp_i]}{[qr] [qp_i]} 
\label{sumnotcol}
\end{align}
where $Q_i = 1$ for a particle or $Q_i=-1$ for an antiparticle (incoming particles would get a relative minus sign).
Next,
define new massless four-vector $t^\mu_j$.
We can use the Schouten identity to write
\be
[r p_i][q t_j] = [qp_i][r t_j] + [t_jp_i][q r]
\ee
This holds for any $i$, so in particular,
\be
\sum_{i} Q_i 
\frac{[rp_i]}{[qr] [qp_i]}  = 
\sum_{i} Q_i 
\frac{[r t_j]}{[qr] [q t_j]} +
\sum_{i} Q_i 
\frac{[ t_j p_i]}{[q t_j] [qp_i]} 
\ee
The first sum has no $p_i$ dependence in the spinor products and vanishes by charge conservation, $\sum_i Q_i=0$. 
Thus we have
\be
\bra{p_1\cdots p_N;q} \cO \ket{0} = -g \sqrt{2}
\sum_{i} Q_i 
\frac{[ t_j p_i]}{[q t_j] [qp_i]}  \label{nor}
\ee
which is explicitly $r$-independent.

Now let us take $t_j$ to point in a direction not collinear to $p_j$. Then all the spinor products which appear in  \Eq{nor}
are $\cO(\lambda^0)$ except for $[q p_j] \sim \lambda$. Thus, at leading power, using $Q_j=1$, we find
\be
\bra{p_1\cdots p_N;q} \cO \ket{0} \LPeq
-g \sqrt{2}  \frac{[ t_j p_j]}{[q t_j] [qp_j]} 
\label{tform}
\ee
Note that $t_j$ can be thought of as an example of a generic-$r$ reference vector. In particular, taking $r=t_j$ in \Eq{genRoneemit} gives exactly \Eq{tform}.
In collinear-$r$, where $r=p_j$, the self-collinear-emission from the $p_j$ line is exactly zero and \Eq{tform}
is produced from the sum of all {\emph{other}} diagrams. 
In fact, in  collinear-$r$, \Eq{tform} can be written as
\be
-g\,\sqrt{2} 
\frac{[ t_j p_j]}{[q t_j] [qp_j]} 
	\overset{p_j=r}=
	-g\,\sqrt{2}  \frac{[t_j r ]}{[q t_j][q r ]}=
g\,\frac{t_j\cdot\epsilon^-_q}{t_j\cdot q}
\ee
which is exactly the amplitude coming from a Wilson line in the $t_j$ direction, as in \Eq{eikvert}, but with opposite sign because \Eq{eikvert} involves the conjugated Wilson line.

The above analysis motivates improving  \Eq{matelmfactsQED} by adding Wilson lines in the $t_j$ directions:
\be
\l p_1 \cdots p_N; q_{a_1} \cdots q_{b_N} | \,\cO\, | 0 \r
\LPeq
\bra{p_1;q_{a_1} \cdots q_{b_1}}\phi^* W_1 \ket{0} \dots
\bra{p_N;q_{a_N} \cdots q_{b_N}} W_N^\dg \phi_N \ket{0}
\label{collphotfact}
\ee 
where the outgoing Abelian Wilson line is
\be
W_j(x) = \exp\bigg( -ig \int_0^\infty ds\,  t_j\cdot A(x^\mu + s\,t_j^\mu)\, e^{-\varepsilon s}\bigg)
\ee
To check \Eq{collphotfact}, we evaluate one of the terms on the right-hand-side with one emission.
This emission can come out of $\phi$ or out of $W_j$, giving
\be
\bra{p_j ;q }\phi^* W_j \ket{0} = -g\,\sqrt{2}  \frac{[r p_j ]}{[q r][q p_j ]}-g\,\sqrt{2}  \frac{[t_j r ]}{[q t_j][q r ]}\\
\ee
Using the Shouten identity this simplifies to
\be
\bra{p_j ;q }\phi^* W_j \ket{0} =
-g \sqrt{2}  \frac{[ t_j p_j]}{[q t_j] [qp_j]} 
\ee
Thus the factorized expression is $r$-independent and agrees with  \Eq{tform}, which is the full matrix element at leading power.

More generally, the factorized expression will be $r$-independent for any number of emissions since the operators $\phi^* W_j$ in the
matrix elements are gauge invariant. Moreover, since the Wilson line contributions to the matrix element are of the form
$ \frac{[t_j r ]}{[q t_j][q r ]}$ which scale like $\lambda^0$ in generic-$r$, they can be set to 1 in generic-$r$. Thus 
\Eq{collphotfact} reduces to \Eq{matelmfactsQED} which we have already shown agrees with full scalar QED at
leading power in generic-$r$. Since \Eq{collphotfact} is reference-vector independent and agrees with full scalar QED for a specific reference-vector choice, it must agree for all reference vectors. Hence, we have proven the factorization formula in \Eq{collphotfact}.

In summary, we have shown that matrix elements of the gauge-invariant operator, $\cO = \frac{1}{(N/2)!}|\phi|^N$, in states with $N$ collinear sectors, each composed of a collinear scalar and many collinear photons, factorize into $N$ separately-gauge-invariant matrix elements as in \Eq{collphotfact}.
We can write collinear factorization succinctly as
\be
\bra{X_1 \cdots X_N}  \cO \ket{0} 
\LPeq
\bra{X_1}\phi^* W_1 \ket{0} \cdots
\bra{X_N} W_N^\dg \phi \ket{0}
\label{collfactgen}
\ee
This expression holds for any choice of polarization vectors. In fact, it holds for any hard-scattering operator $\cO$ even with additional derivatives in it or for scattering mediated by operators with different numbers of fields. The effect of considering more general scattering is to simply multiply \Eq{collfactgen} by an overall function, $H(P_1,\ldots,P_N)$, where $P_i$ is the total momentum of the $i$-th sector, as discussed in 
Section~\ref{sec:IdentPart}.

In proving the above factorization of collinear sectors, we used states of one charged scalar and an arbitrary number of photons collinear to each direction. This was done for simplicity; the factorization holds for more general states, $ \ket{X_j} $, of an arbitrary number of $j$-collinear scalars and photons that carry the quantum numbers of a single scalar.
The splitting of a photon into particle-anti-particle pair is reference vector independent and therefore only enhanced for self-collinear emissions and splittings.  Thus, the diagrammatic factorization of \Eq{diagfactsQED} becomes (in generic-$r$)
\be
\sum\quad \fd{3.5cm}{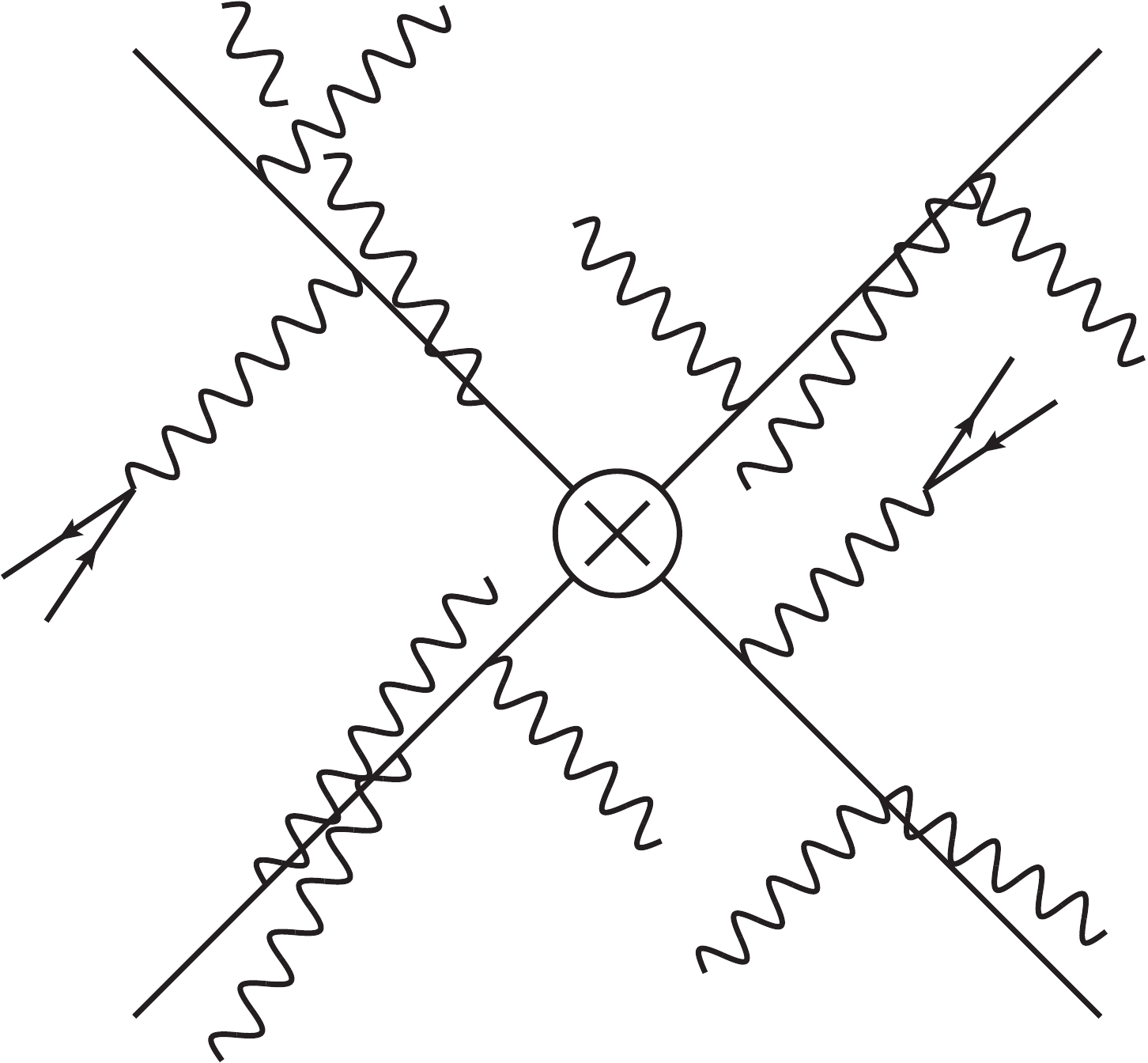} 	\quad\LPeq\quad
\sum\quad \fd{3.3cm}{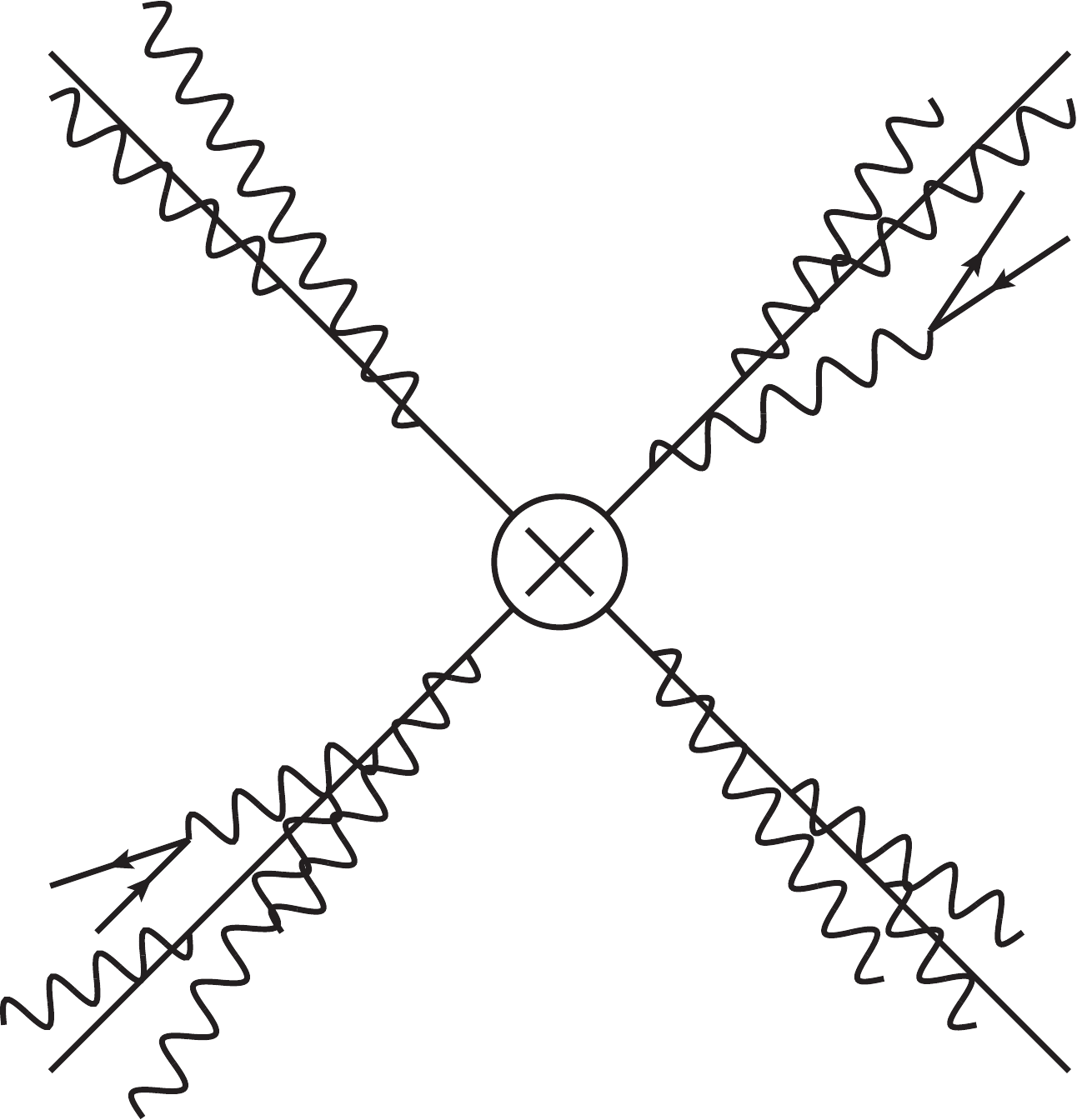} 
\label{diagfactsQED_full}
\ee
and the rest follows exactly as above.

\subsection{Soft Factorization}
\label{sec:sQEDSF}

In this subsection we will ignore any collinear dynamics by considering states with only one 
collinear momentum in each sector.
 We discuss the simultaneous factorization of the soft and collinear sectors in Section~\ref{sec:sQEDSSCF}. In terms of matrix elements, we will consider the simplified problem of factorizing the soft emissions in
\be
\bra{p_1\cdots p_N;X_s} \cO \ket{0}
\ee
from the hard-scattering matrix element $\bra{p_1\cdots p_N} \cO \ket{0}=1$.
As a further initial simplification, we will consider $\bra{X_s}$ to consist of soft photons only. Any soft scalars in $\bra{X_s}$ must couple to collinear lines through a virtual-soft photon, so we can deal with soft scalars once we have understood how soft photons decouple.

A convenient feature of the soft limit is that we do not need to worry about polarization-vector subtleties since soft emissions are not associated with a specific direction. Indeed, as observed in Section~\ref{sec:spinorhelicity},
we can just use $k\sim \lambda^2$ and $\epsilon\sim1$.
Alternatively, we can choose generic-$r$ for all of the soft polarizations; we will
 see that our final result is gauge-invariant and hence independent of this choice. 
 
 Now, consider the addition of soft emissions to the hard-scattering matrix element. For one emission, the graph is 
\be
\fd{3.5cm}{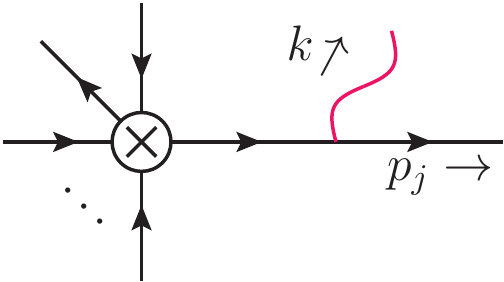} = -g\,\frac{ p_j\cdot\epsilon_k}{p_j\cdot k \pie}\, 
	\sim \frac{g}{\lambda^2}\,
		\label{sQEDsoft1}
\ee
where soft photons are colored red and have long wavelengths. In scalar QED there are also diagrams with a 4-point vertex:
\be 		
\fd{3.5cm}{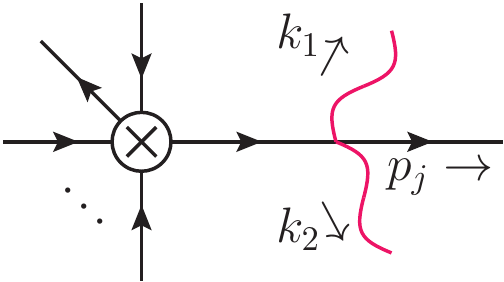} = \frac{-g^2\,\epsilon_{k_1}\cdot\epsilon_{k_2}}{p\cdot(k_1+k_2)\pie} \,
 \sim \frac{g^2}{\lambda^2}\, 
		\label{sQEDsoft2}
\ee
We see that when considering soft emissions, diagrams involving the four-point vertex (which scale like $\lambda^{-2}$) are subleading
compared to diagrams involving two emissions from three-point vertices (which scale like
$\lambda^{-4}$).

Next, consider the sum of the most-enhanced soft emissions off of a single collinear line. 
For $\ell$ soft emissions off of a scalar with momentum $p_j^\mu = E_j n_j^\mu$,
we can use $p_j + k_i \LPeq p_j$ to write 
the matrix element as
\be
\sum_\text{perms} \;\fd{6.9cm}{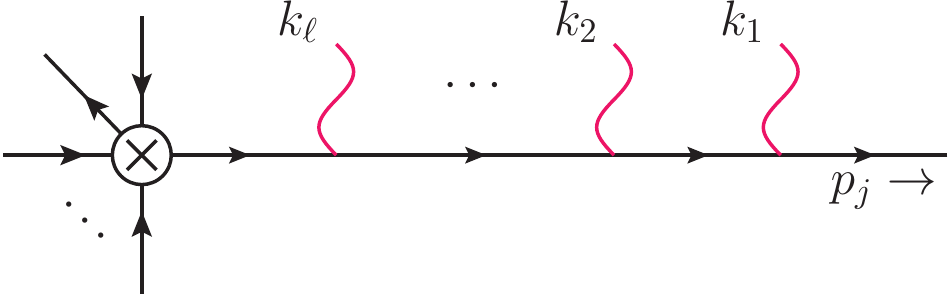}
  \LPeq \sum_\text{perms}(-g)^\ell \, \frac{p_j\cdot \epsilon_1}{p_j\cdot k_1}\frac{p_j\cdot \epsilon_2}{p_j\cdot (k_1+k_2)}\cdots\frac{p_j\cdot \epsilon_\ell}{p_j\cdot \sum_{i=1}^\ell k_i} 
 \ee
 where the sum on ``perms'' means to sum over all permutations of the soft photons.
Note that each term is independent of the energy of the scalar, $E_j$ and only depends
on its direction $n_j^\mu$.
After some algebra (known as the eikonal identity), this reduces to
\be
\sum_\text{perms} \;\fd{6.9cm}{paperfigs/softcolldec1_vert.pdf}
\;\LPeq\;
(-g)^\ell \, \frac{n_j\cdot \epsilon_1}{n_j\cdot k_1}\frac{n_j\cdot \epsilon_2}{n_j\cdot k_2}\cdots\frac{n_j\cdot \epsilon_\ell}{n_j\cdot k_\ell} 
\label{softemisqline}
\ee
This form of the amplitude indicates that in the soft limit, the separate soft emissions are totally uncorrelated in scalar QED. Moreover, each factor can immediately be seen to be reproducible as the matrix element of a Wilson line, as in \Eq{eikvert}. In the Abelian case, the Wilson line is
\be
Y_j^\dg(x) = \exp\bigg( ig \int_0^\infty ds\, n_j\cdot A(x^\mu + s\, n_j^\mu)\, e^{-\varepsilon s}\bigg)
\ee
and the result is
\be
\sum_\text{perms} \;\fd{6.9cm}{paperfigs/softcolldec1_vert.pdf} 
\;\LPeq\; \bra{k_1\cdots k_\ell} Y^\dg_j(0) \ket{0}
\label{keyeqSF}
\ee
Although the delightfully simple form in~\Eq{softemisqline} is particular to Abelian gauge theories (it is indicative of Abelian exponentiation\cite{Yennie:1961ad}),
 that multiple soft emissions can be written in terms of matrix elements of Wilson lines, as in \Eq{keyeqSF} is also true in the non-Abelian case, as we discussion in Section~\ref{sec:QCD}.

The generalization of \Eq{keyeqSF} to soft emissions off of multiple lines simply requires the inclusion of multiple Wilson lines on the right-hand-side. In terms of operator matrix elements, the general result is
\be
\bra{p_1\cdots p_N; k_1\cdots k_\ell} \cO \ket{0}
	\;\LPeq\; \bra{p_1\cdots p_N} \cO \ket{0}\, \bra{k_1\cdots k_\ell} Y_1^\dg\cdots Y_N \ket{0} 
\label{softphotfactsQED}
\ee
We will now prove this using only \Eq{keyeqSF} and a straightforward enumeration of the diagrams associated with the contractions on the two sides. Because our proof only uses  \Eq{keyeqSF} we will be able to recycle it for soft-collinear factorization, spinor QED and QCD below.

First, note that both sides of \Eq{softphotfactsQED}, in the soft limit, consist of a sum of terms with different numbers of $\frac{n_i\cdot \epsilon_k}{n_i\cdot k}$ factors for each $i$. On the left-hand side, each factor comes from the contraction of a photon with the $i$-th scalar and on the right-hand side, from a contraction of a photon with the $i$-th Wilson line. Thus, let us define the set of integers $\{ \ell_i\}$ corresponding to a particular partitioning of the number of photons connecting to each direction. That is, $\ell_1$ photons connect to the first scalar, $\ell_2$ to the second, and so on.  It is then clear that \Eq{softphotfactsQED} should hold for each set $\{ \ell_i\}$ separately.

For each Feynman diagram, it is easy to read off what $\{ \ell_i\}$ is.
Let $\cD_{\{ \ell_i\}}$ be some diagram with $\ell_i$ photons attached to the $i$-th leg.
For example, we might take 
\be
\cD_{1,2,2,1,3}  \;\equiv\; \fd{3.5cm}{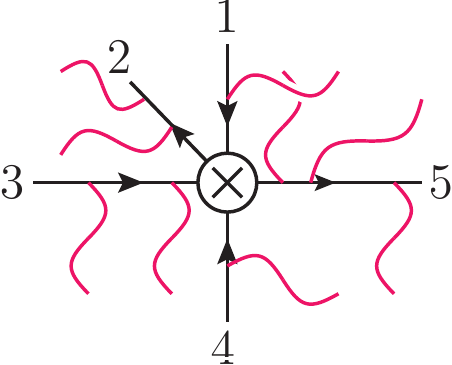}
\ee
If we choose a fiducial diagram $\cD_{\{\ell_i\}}$ for each possible set of integers $\{ \ell_i\}$ satisfying $0\leq \ell_i \leq \ell$ and $\sum_i \ell_i = \ell$, then  
we can write the matrix element as
\be
\bra{p_1\cdots p_N;k_1\cdots k_\ell} \cO \ket{0} 
=
 \sum_{\{ \ell_i\}}
~
\sum_{\substack{ \text{perms of} \\  \kperm } }
\sum_{\substack{\text{perms} \\ \text{on}~p_1}}
\cdots
\sum_{\substack{\text{perms} \\ \text{on}~p_N}}
  \cP \big[ \cD_{\{ \ell_i\}} \big]
\label{toomanysums}
\ee
The first sum is over the partitionings   $\{ \ell_i\}$. The second sum, denoted ``perms of $\kperm$'',
is over the permutations of which $\ell_i$ photons, $k^i_1 \cdots k^i_{\ell_i}$, connect to which leg, for each $i$.
The ``perms on $p_j$'' changes the ordering by which the photons connect to the $j$-th line, keeping  $\{ \ell_i\}$ and $\kperm$ fixed.
 Finally,   $\cP [ \cD_{\{ \ell_i\}} ]$ means apply the product of all these permutations to the fiducial diagram for the given   $\{ \ell_i\}$.

 Now, let us evaluate these sums. For the ``perms of $p_N$'' sum, we are to hold the emissions off of legs $1$ through $N-1$ fixed and sum only over permutations of the photons attached to the $N$ leg.
 This rest-of-the-diagram
 provides an overall multiplicative factor to \Eq{keyeqSF} which has no effect on the correspondence
 between the soft photons attached to leg $N$ and the Wilson line, $Y_N$. Thus we have
\be
\sum_{\overset{\text{perms}}{\text{of}~p_N}}
  \cP \big[ \cD_{\ell_1,\ldots,\ell_{N-1}, \ell_N} \big]
=
	\bra{k^N_1\cdots k^N_{\ell_N}} Y_N \ket{0}
 \;	\cD_{\ell_1, \ldots, \ell_{N-1},0}
\label{DellD0}
\ee
 Similarly, we can now proceed to the evaluation of the
 ``perms of $p_{N-1}$'' sum to produce a matrix element of the $N-1$ Wilson line.  We can continue in this way until there are no lines left and we have only  $\cD_{0,\ldots,0}= \bra{p_1\cdots p_N} \cO \ket{0}$. Thus
  \Eq{toomanysums} becomes
\be
\bra{p_1\cdots p_N;k_1\cdots k_\ell} \cO \ket{0} \LPeq \bra{p_1\cdots p_N} \cO \ket{0} \;
 \sum_{\{ \ell_i\}}
~
\sum_{\overset{ \text{perms of} } { \kperm } }
\prod_{i=1}^N\bra{k^i_1\cdots k^i_{\ell_i}} Y^\dg_i \text{ or } Y_i \ket{0}
\label{sumsumprod}
\ee
where ``$Y^\dg_i \text{ or } Y_i$'' means $Y^\dg_i$ if $p_i$ is a particle or $Y_i$ if $p_i$ is an anti-particle.

The last step is simply to note that  
$\sum_{\{ \ell_i\}} \sum_{ \kperm }$
exactly coincides with the sum of contractions of the composite operator $Y_1^\dg \cdots Y_N$ with the photons in the external state $\bra{k_1\cdots k_\ell}$. This should not come as a surprise since these sums came from the contractions of the Lagrangian insertions with the same external state on the left-hand side of \Eq{toomanysums}. Hence, \Eq{sumsumprod} reduces to \Eq{softphotfactsQED} which was our desired result.

Now that we have understood how soft photons decouple from collinear lines, we can immediately generalize \Eq{softphotfactsQED} to include soft scalar-anti-scalar pairs in the state $\bra{X_s}$
which leads to
\be
\bra{p_1\cdots p_N; X_s} \cO \ket{0}
	\LPeq \bra{p_1\cdots p_N} \cO \ket{0}\, \bra{X_s} Y_1^\dg\cdots Y_N \ket{0} 
\label{softfactsQED}
\ee
This generalization is immediate  because soft scalars must couple through a soft photon which is approximately on-shell and we have shown that the latter couples to collinear lines via the soft Wilson lines, $Y_j$. Therefore, if we simply let $\bra{X_s} Y_1^\dg\cdots Y_N \ket{0}$ evolve under the full scalar QED Lagrangian, we can describe soft scalar production by an emission from a soft Wilson line, followed by a splitting from the Lagrangian. A similar story holds for pair creation from a collinear photon and was discussed at the end of Section~\ref{sec:sQEDCF}.

The bright side of the heavy notation that we introduced in this section is that every equation after \Eq{keyeqSF} only relied on how fields are contracted with states in quantum field theory. Therefore, as long as \Eq{keyeqSF} continues to hold, the above proof will work for path-ordered Wilson lines in QCD as well as with any modification to $\cD_{0,\ldots,0}$. The former will be used in Section~\ref{sec:QCD} to show soft factorization in QCD and the latter will be used in the next section to show soft-collinear factorization in scalar QED

\subsection{Simultaneous Soft-Collinear Factorization}
\label{sec:sQEDSSCF}

We now tackle the full problem of factorizing matrix elements of the form
\be
\bra{X_1\cdots X_N;X_s} \cO \ket{0}
\ee
where $\bra{X_j}$ is a state of many collinear particles,
 $\bra{X_s}$ is a state of many soft particles,
  and $\cO = \frac{1}{(N/2)!}|\phi|^N$ is the hard-scattering operator. 
As above, since soft scalars must come from pair creation initiated by a soft photon,
when discussing how the soft sector decouples from the collinear sectors, we only need to worry about soft photons.
We therefore assume $\bra{X_s}$ just contains soft photons.

\subsubsection{Soft coherence \label{sec:coherence} }
One way to understand why soft-collinear factorization holds is to think of it in terms of coherence~\cite{Catani:1999ss,Ermolaev:1981cm}.
In a classical theory, the electromagnetic field far away from a set of charged particles is only
sensitive to the net charge at leading order in the multipole expansion. In the same way, soft
radiation is only sensitive to the net charge of a set of  particles all collinear to the same direction. 
For example, if there are two particles with charges $Q_1$ and $Q_2$ and momenta
$p_1$ and $p_2$ with $p_1 \parallel p_2 \parallel n$, then the two diagrams for emitting a
soft photon of momentum $k$ and polarization $\epsilon$ add as follows:
\begin{align}
&\fd{5cm}{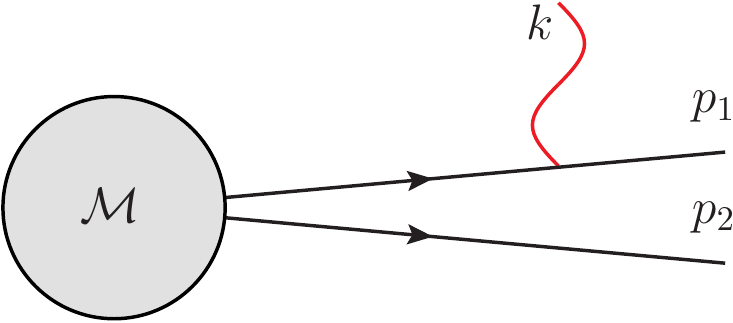} \quad+\quad \fd{5cm}{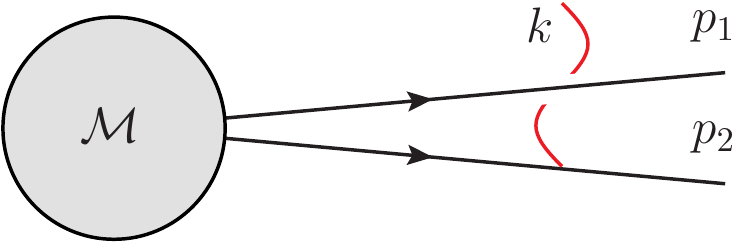} \notag
\\& \hspace{1cm} = 
\bigg( Q_1 \frac{p_1 \cdot \epsilon} { p_1 \cdot k} +Q_2 \frac{p_2 \cdot \epsilon} { p_2 \cdot k} \bigg)\times\cM
\;\LPeq\; (Q_1+Q_2) \frac{n \cdot \epsilon} { n\cdot k}\times\cM
\label{softcoher}
\end{align}
This amplitude is the same as one where the soft photon was emitted from a single line in the $n$ direction with charge $Q_1+Q_2$. That is the general idea, at least.

Unfortunately, \Eq{softcoher} does not hold generally.
The problem is that
 $\cM$ could depend on $k$ differently in the two diagrams, in which case we would not be able to
simply take $k\to 0$ because $k\cdot p_1$ and $k\cdot p_2$ are the same size as $p_1\cdot p_2$, namely $\cO(\lambda^2)$.
In other words, we must worry about diagrams that ``tangle'' the soft and collinear emissions. For example, with one soft and one collinear photon, there are three diagrams in scalar QED which
connect these photons to the same leg:
\begin{align}
&\fd{4.5cm}{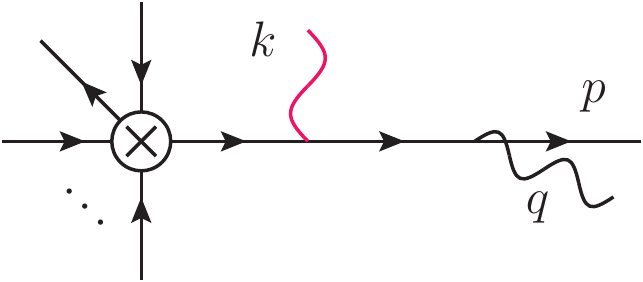} 
= \frac{-g\,(p + q)\cdot\epsilon_{k}}{p\cdot q + (p+q)\cdot k}\, \frac{-g\, p\cdot\epsilon_{q}}{p\cdot q}
	\sim
\frac{-g\,(p + q)\cdot\epsilon_{k}}{\lambda^2 + \lambda^2}\, \frac{-g\, p\cdot\epsilon_{q}}{\lambda^2}
\notag\\ &
\fd{4.5cm}{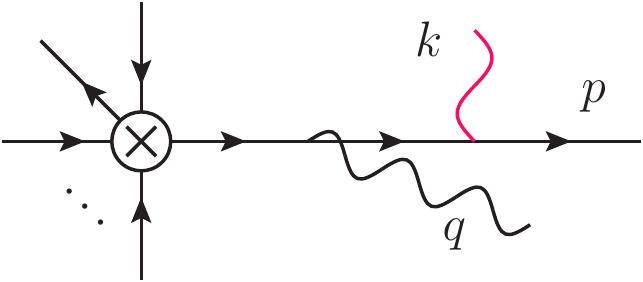} 
= \frac{-g\,(p+k)\cdot\epsilon_{q}}{p\cdot q + (p+q)\cdot k}\, \frac{-g\,p\cdot\epsilon_{k}}{p\cdot k}
	\sim 
\frac{-g\,p\cdot\epsilon_{q}}{\lambda^2 + \lambda^2}\, \frac{-g\,p\cdot\epsilon_{k}}{\lambda^2}
\label{tanglediags}
 \end{align}
 and 
 \be
 \fd{4.5cm}{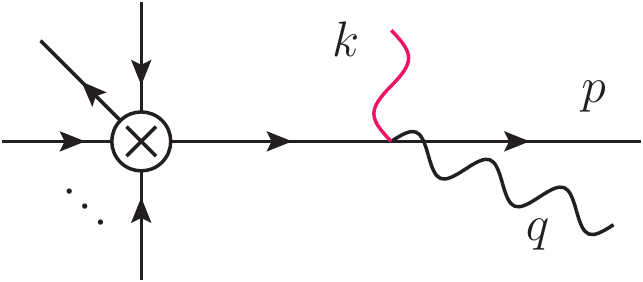} 
= \frac{-g^2\,\epsilon_k\cdot\epsilon_{q}}{p\cdot q + (p+q)\cdot k}
	\sim 
 \frac{-g^2\,\epsilon_k\cdot\epsilon_{q}}{\lambda^2 + \lambda^2}  \notag
 \ee
Here, the soft photons are colored red and are drawn with broader wiggles than the collinear photons. The naive soft coherence argument of \Eq{softcoher} would imply that only the second diagram should contribute, but clearly the first diagram is the same order in the power-counting.

Of course, coherence does actually hold, and it is not too hard to simplify these amplitudes to see it
directly. With spinor-helicity methods, we can prove some useful and non-obvious identities, such as
\begin{align}
\epsilon^+_q \cdot \epsilon^+_k &= 	\frac{p\cdot\epsilon^+_q\, q\cdot \epsilon^+_k}{p\cdot q} 
							   + 	\frac{k\cdot\epsilon^+_q\,p\cdot\epsilon^+_k}{p\cdot k} 
							   -  	\frac{p\cdot\epsilon^+_q\,p\cdot\epsilon^+_k q\cdot k}{p\cdot q\,p\cdot k} \\
\epsilon^+_q \cdot \epsilon^-_k &= 	\frac{p\cdot\epsilon^+_q\, q\cdot \epsilon^-_k}{p\cdot q} 
							   + 	\frac{k\cdot\epsilon^+_q\,p\cdot\epsilon^-_k}{p\cdot k} 
							   -  	\frac{p\cdot\epsilon^+_q\,p\cdot\epsilon^-_k q\cdot k}{p\cdot q\,p\cdot k}
							   - \underbrace{\frac{\l q p\r [kp]}{[pq]\l pk\r}}_{=1}
\end{align}
where both of these equations hold for any reference vector choices and any four-momentum $p$, and the last term is 1 if the momenta are real. The other possible helicity choices can be found be conjugating the above equations. Now we can simplify the tangled diagrams of \Eq{tanglediags} as follows:
\begin{align}
&	\fd{4cm}{paperfigs/softcollambi1_vert.pdf} 
\;+\; 	\fd{4cm}{paperfigs/softcollambi2_vert.pdf} 
\;+\; 	\fd{4cm}{paperfigs/softcollambi3_vert.pdf} 
\\&\,=	\frac{(-g)^2}{p\cdot q + (p+q)\cdot k} \bigg[
		(p + q)\cdot\epsilon_{k}\, \frac{p\cdot\epsilon_{q}}{p\cdot q}
      +	(p+k)\cdot\epsilon_{q}\, \frac{p\cdot\epsilon_{k}}{p\cdot k}
      -		\epsilon_k\cdot\epsilon_{q} \bigg]
\\&=\begin{dcases}  
(-g) \frac{p\cdot\epsilon^\pm_{q}}{p\cdot q}\, (-g)\frac{p\cdot\epsilon^\pm_{k}}{p\cdot k}	 
			& \quad\text{for }\pm\pm\text{ polarizations}\\ 
(-g) \frac{p\cdot\epsilon^\pm_{q}}{p\cdot q}\, (-g)\frac{p\cdot\epsilon^\mp_{k}}{p\cdot k}
	+ 	\frac{(-g)^2}{p\cdot q + (p+q)\cdot k} & \quad\text{for }\pm\mp\text{ polarizations}
\end{dcases}
\label{twotangelcases}
\end{align}
where this equality is completely general; it holds for any reference vector choice for either photon as well as any on-shell four momenta, $p,q,k$.

\Eq{twotangelcases} 
 says that the sum of the three tangled diagrams reduces to an eikonal form plus a term that is polarization vector independent. Since the leading power diagrams for one soft and one collinear emission scale like $g^2/\lambda^3$, the extra term in \Eq{twotangelcases} is a power correction.
Thus, at leading power the sum of the tangled graphs reduces to the eikonal
form, which is simply the product of the separate amplitudes for soft and collinear emissions. In particular, the soft photon factorizes
off as expected, and is only sensitive to the net charge of the scalar, independent of whether there
are collinear photons nearby.

Although it is surely possible, it would certainly be cumbersome to evaluate matrix elements explicitly for an arbitrary number of soft and collinear emissions in scalar QED. Moreover, analyzing the diagrams directly in scalar QED would also not easily generalize to an analysis for QCD.
Fortunately, soft-collinear factorization can be derived much more simply by exploiting reference-vector independence, which generalizes easily to more-complicated gauge theories.

\subsubsection{General soft-collinear factorization}
\label{sec:sQEDGSCF}

We begin with a lemma:
\newtheorem*{thm}{Lemma}
\begin{thm}
Two expressions that are independent of the choice of reference vectors, $r_i$, and that agree at leading power for particular $r_i$, must agree at leading power for any $r_i$.
\end{thm}
\noindent
This lemma is the trivial statement that two constant functions that agree somewhere, agree everywhere.
It is nevertheless extremely powerful. Since the matrix elements in full gauge theories and factorized expressions
in terms of Wilson lines are both $r_i$-independent, this lemma reduces the problem
of proving factorization to working with particular choices of $r_i$.

Consider first states $\bra{p_1\cdots X_j \cdots p_n; X_s}$ with an arbitrary number of soft photons, but where all of the collinear photons are collinear to the same direction $n^\mu_j$. 
Let $r_s$  denote the reference vector for the soft photons and $r_c$ denote the reference vector for the collinear photons.
First choose generic-$r$ for the collinear photons so that the relevant diagrams at leading power have only self-collinear emissions. Now, since the soft momenta $\{k\}$ are not collinear to $p_j$,
choosing $r_s = p_j$ does not make their polarization vectors, $\epsilon_k$, 
enhanced.\footnote{The region of phase-space for which $k$ is both collinear to $p$ and soft
is not interesting at tree-level.  At tree-level, one is free to choose whether to call this photon soft or collinear; it will factorize
either way. At loop-level, the soft-collinear region is more subtle~\cite{Manohar:2006nz}.}
 With these choices, the diagrams which contribute at leading
power have the collinear
photons coming off  \emph{only} the $j$-th scalar and the soft photons coming off of \emph{all but} the $j$-th scalar:
\be
\bra{p_1\cdots X_j \cdots p_N ; X_s} \cO \ket{0} 
 \overset{r_s=p_j}{\overset{\text{gen}-r_c}{\LPeq}}
 \sum_{\text{coll: only }j} \;\; \sum_{\text{soft: no }j}\; 
 \fd{5.3cm}{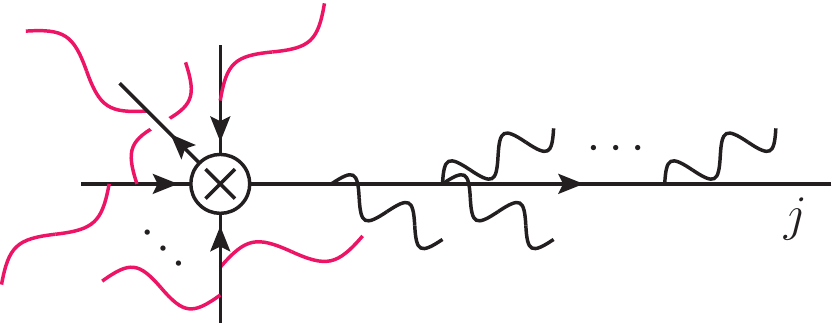}
 \label{GSCFstart}
 \ee 
 Since no diagram in this set has soft and collinear photons connected to the same line, we can use
 the separate arguments for soft and collinear factorization to show that these diagrams factorize. That is, writing the sum over soft emissions as in \Eq{toomanysums}, using \Eq{DellD0} and then recombining the leftover sums, we have
 \be
 \bra{p_1\cdots X_j \cdots p_N ; X_s} \cO \ket{0} 
  \overset{r_s=p_j}{\overset{\text{gen}-r_c}{\LPeq}}
  \bra{p_1\cdots X_j \cdots p_N} \cO \ket{0} 
  \times
\bra{X_s} Y_1^\dag \cdots Y_{j-1} Y_{j+1} \cdots Y_N\ket{0}
\label{manysoftsnoj}
\ee
Now, since we are assuming $r_s=p_j$, any matrix element of $Y_j$ with soft photons is power suppressed: $\bra{X_s} Y_j \ket{0} \sim 0$. Therefore, we can add $Y_j$ to the right-hand side of \Eq{manysoftsnoj} without changing it at leading power. Thus, we have 
\be
\bra{p_1\cdots X_j \cdots p_N ; X_s} \cO \ket{0} 
\LPeq
\bra{p_1\cdots X_j \cdots p_N} \cO \ket{0} 
\bra{X_s} Y_1^\dagger\cdots Y_N \ket{0}
\label{pXp}
\ee
Although we have only shown that this holds when $r_c$ is generic and $r_s = p_j$,
by the lemma, since both sides are reference-vector independent, it must hold
for any reference vectors.

There is not a diagram-by-diagram correspondence in \Eq{pXp}. However,
it is possible to identify sets of diagrams whose sums agree at leading power. 
We will continue to work in generic-$r$  for the collinear photons, so that all the diagrams
which contribute at leading power to \Eq{pXp} have the collinear photons connected
to leg $j$. Then, the particular sets of diagrams we need  in order to prove soft-collinear
factorization are the equivalent of \Eq{keyeqSF}, namely, diagrams where all the soft momenta connect to the $j$ line:
\be	
\label{softpermuted}
\sum_{\substack{\text{ perms of} \\ \text{soft on}~p_j}}
\fd{5.3cm}{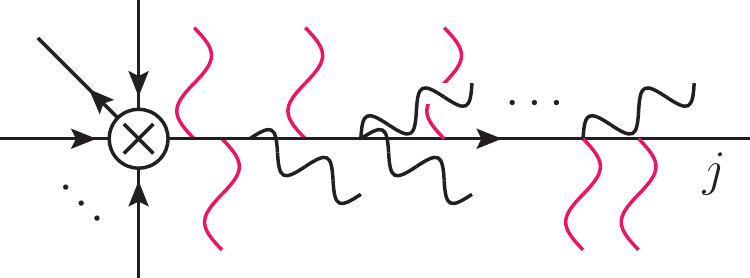} 
	\;\LPeq\;
\fd{5.3cm}{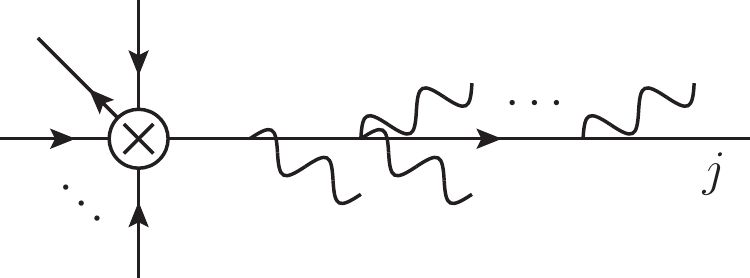}
 	\times  \bra{X_s}  Y_j^\dg  \ket{0}
\ee
Here, the sum is over permutations where the soft photons connect, holding the topology of the
collinear photons fixed but arbitrary.
This diagrammatic relation is the key to soft-collinear factorization: it says that soft photons can be simply stripped off of Feynman diagrams, like leaves off a sprig of thyme.

We will prove \Eq{softpermuted} by induction on the number of soft photons.
For zero soft-photons, the equation is trivially satisfied. So let us assume \Eq{softpermuted} holds for
any number of soft photons less than $n$. 
Now, consider
the diagrams which contribute to the left-hand side of \Eq{pXp} with $n$ photons and let $\cD_{\{\ell_i\}}$ be a fiducial diagram in that sum with $\ell_i$ soft photons on the $i$-th collinear line and $\sum_i\ell_i = n$. Note that, here  $\cD_{\{\ell_i\}}$ has a fixed topology of self-collinear emissions in the $j$-th sector as well as all of the soft photons. Now, as in \Eq{toomanysums}, write the sum of diagrams in \Eq{pXp} as:
\be
\bra{p_1\cdots X_j \cdots p_N;X_s} \cO \ket{0} =
\sum_{\substack{ \text{perms of} \\ \text{coll.} } }
\sum_{\{ \ell_i\}} 
\sum_{\substack{ \text{perms of} \\ \kperm } }
\sum_{\substack{\text{perms} \\ \text{on}~p_j}}
\cdots
\sum_{\substack{\text{perms} \\ \text{on}~p_N}}
  \cP \big[ \cD_{\{ \ell_i\}} \big]
\label{toomanysumsXj}
\ee
For simplicity, we sum over the permutations of attachments to the $j$-th line last. The sums with $i \neq j$ can be performed using \Eq{DellD0},
which holds even with collinear emissions on the $j$ leg. The last sum to do is over the ``perms on $p_j$'' of $\cD_{0,\ldots,\ell_j,\ldots,0}$.
If $\ell_j <n$, we can perform this sum using the induction hypothesis:
\be
\sum_{\substack{\text{perms} \\ \text{of}~p_j }}
  \cP \big[ \cD_{0,\ldots,\ell_j,\ldots,0} 
\big]
\LPeq
\cD_{0,\ldots,0}
\; \bra{k^j_1\cdots k^j_{\ell_j}} Y^\dg_j \ket{0}
 \For \ell_j < n
\label{indhyp}
\ee
The only sum we cannot perform is the one when all $n$ photons attach to leg $j$. 
So let us add and subtract $\sum_{ \text{perms of coll.}} \cD_{0,\ldots,0} \bra{k_1\cdots k_n} Y^\dg_j \ket{0}$
to \Eq{indhyp}. When we add it, we have a sum of
terms with matrix elements of soft Wilson lines just like \Eq{sumsumprod}. These Wilson lines can then combine into a single composite operator.
We thus have
\begin{multline}
\bra{p_1\cdots X_j \cdots p_N;X_s} \cO \ket{0} \LPeq 
\bra{p_1\cdots X_j \cdots p_N} \cO \ket{0}  \bra{X_s} Y_1^\dagger\cdots Y_N \ket{0} \\
+ \sum_{\substack{ \text{perms of} \\ \text{coll.} } }
\bigg[
\sum_{\substack{\text{perms} \\ \text{on}~p_j}}\cD_{0,\ldots,n,\ldots,0} 
\;-\;
\cD_{0,\ldots,0}\, \bra{k_1\cdots k_n} Y^\dg_j \ket{0}
\bigg]
\end{multline}
Since the first line holds on its own, by \Eq{pXp}, the second line must vanish. Moreover, since nothing we have said depended on summing over the collinear permutations, the term in square brackets vanishes on its own. This proves \Eq{softpermuted}.

With \Eq{softpermuted} proven, we find ourselves in exactly the position we were in Section~\ref{sec:sQEDSF} with \Eq{softpermuted} taking the place of \Eq{keyeqSF} and $\cD_{0,\ldots,0} $ a specific diagram in $\bra{X_1\cdots X_N} \cO \ket{0}$ instead of being $\bra{p_1\cdots p_N} \cO \ket{0}$. Since the arguments of Section~\ref{sec:sQEDSF} did not depend on $\cD_{0,\ldots,0}$, we can prove that $\bra{X_s} Y_1^\dagger\cdots Y_N \ket{0} $ factors off of each possible collinear diagram in $\bra{X_1\cdots X_N} \cO \ket{0}$ exactly as it was done for $\bra{p_1\cdots p_N} \cO \ket{0}$. Hence,
\be \label{gensoftfact}
\bra{X_1 \cdots X_N; X_s} \cO \ket{0}
  {\overset{\text{gen}-r_c}{\LPeq}}
\bra{X_1 \cdots X_N} \cO \ket{0} \bra{X_s} Y_1 \cdots Y_N^\dag \ket{0}
\ee
Now, since both sides of this equation are $r$-independent, we can drop the restriction 
that the collinear photons are in generic-$r$, by the lemma.

Finally,  since the soft emissions are factorized off, we can now factorize the collinear sectors as in \Eq{collfactgen} and use the same argument as before to allow for soft scalars in $\bra{X_s}$, giving
the final form for factorization in scalar QED:
\be
\begin{array}{|c|}
\hline\\
~~~\bra{X_1\cdots X_N; X_s} \cO \ket{0} 
\LPeq \bra{X_1}\phi^* W_1 \ket{0}
\dots \bra{X_N}W_N^\dg\phi \ket{0}\bra{X_s} Y_1^\dg\cdots Y_N \ket{0}
~~~\\
\\
\hline
\end{array}
\label{genfactsQEDbox}
\ee

Let us review the ingredients that went into this derivation. First, we used separate soft and collinear
factorization. We also used various facts about which classes of diagrams could contribute
with certain reference-vector choices, and of course reference-vector independence. 
We did not use any results specific to scalar QED or even to Abelian gauge theories;
 the identical arguments may be used in any gauge theory to prove soft-collinear factorization.


\subsection{The position-space picture}
\label{sec:physpict}
\begin{figure}[t]
	\begin{center}
\includegraphics[scale=.45]{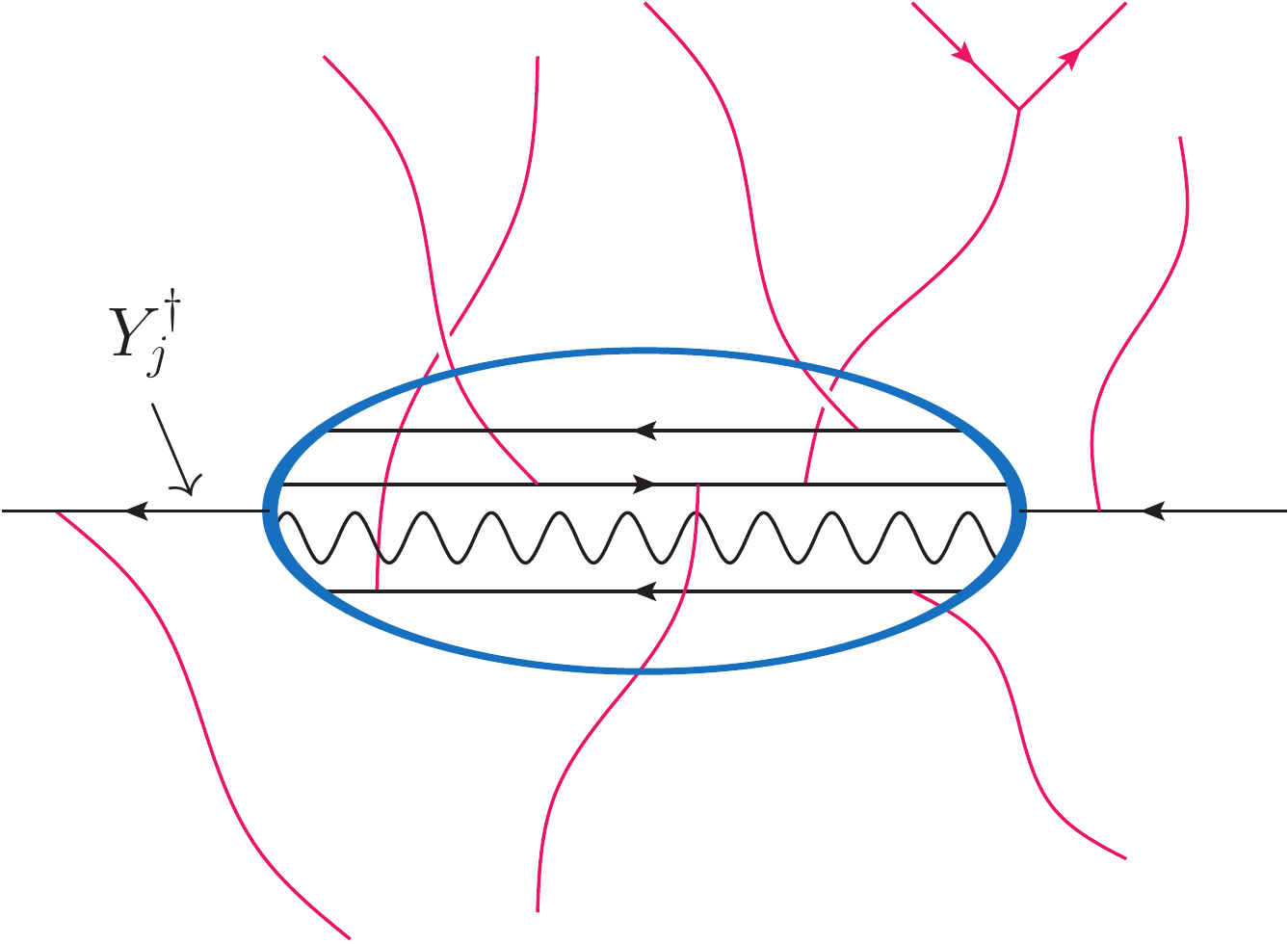}
\caption{The physical picture of soft photons emitted by a classical source, $Y_j^\dg$. The blue oval represents magnification that shows the substructure of the collinear sector which is invisible to the soft photons.}
\label{fig:SoftPhotSources}
	\end{center}
\end{figure}

Before moving on to theories more complicated than scalar QED, it is worth revisiting the physical picture behind factorization since it is identical in spinor QED or QCD. Although matrix elements are rarely
computed in position space, position space is where our physical intuition lies. Since collinear
fields have transverse momenta that scale like $p_\perp \sim \lambda\, Q$, the associated radiation field has a characteristic transverse size of
$ x_\perp \sim (\lambda\,Q)^{-1}$. 
In contrast, since soft momenta scale like $k \sim \lambda^2\, Q$ in all components, the soft radiation
field varies over scales $ x \sim (\lambda\,Q)^{-2}$.  In particular, since soft photons have
wavelengths which are a factor of $\lambda^{-1}$ larger than the width of the collinear sector (the jet), they cannot resolve the jet's substructure, only its net charge. This is shown pictorially in Figure~\ref{fig:SoftPhotSources}.

Outside of the jets, the soft particles interact, and split into scalars on time scales $\delta t \sim (\lambda\,Q)^{-2}$. None of these additional soft particles can probe the jet's substructure either. That is why the soft Lagrangian is a totally decoupled form the collinear Lagrangian -- soft photons only see the jet as a classical source of radiation in the direction $n^\mu$, which is what the Wilson line encodes.
In fact, if one changes to radial coordinates, so the jets become parallel lines 
extending form $\tau = \pm \infty$~\cite{Chien:2011wz},
 one can literally think of the jets as parallel wires whose moving charges
 only leave a collective  imprint on the exterior magnetic field.

The position-space picture of collinear factorization is similar to the soft case just described. Due to the scaling of the momenta of collinear particles, collinear radiation is confined into a set of cones. The radiation in the $j$-th cone (call it the $j$-cone) cannot resolve any of the dynamics outside of that cone; it can only see radiation that is emitted into it. In particular, an individual collinear sector is insensitive to the direction of travel of the charges outside of its cone, just as the soft photons could not resolve the substructure in a collinear sector. Therefore, all the charges outside of the cone can be deformed into the same direction, say $t_j^\mu$, in which case the charges add exactly as in \Eq{softcoher}. This is why an individual collinear sector sees the Wilson line, $W_j$; a classical source of radiation traveling in the $t_j$ direction with opposite charge to that in the cone.

\begin{figure}[t]
	\begin{center}
\includegraphics[scale=.4]{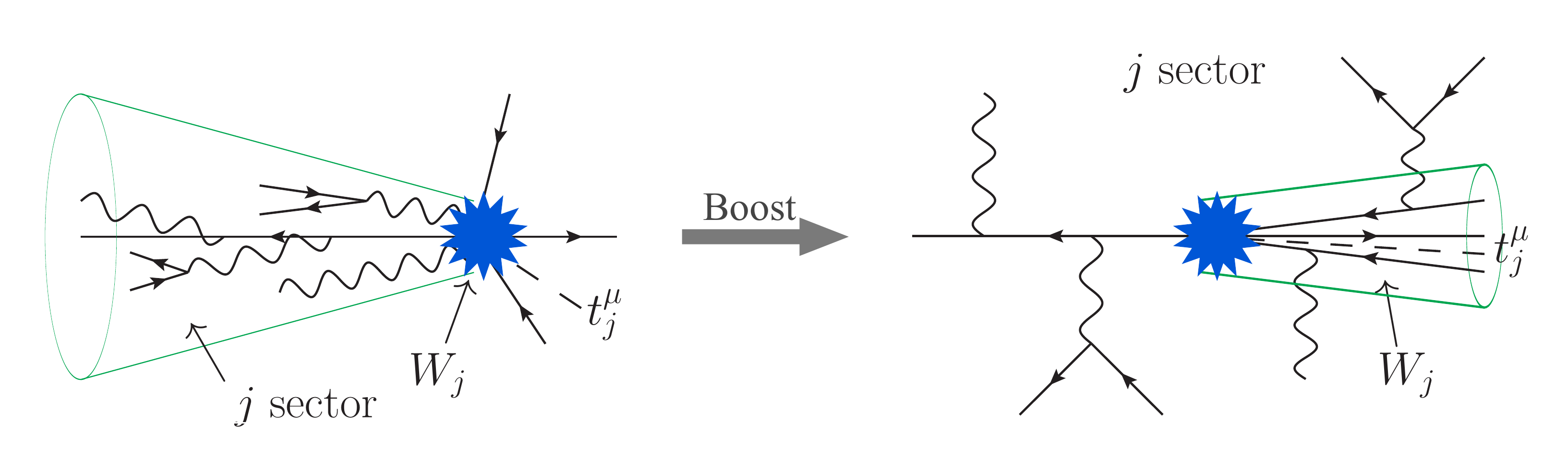}
\caption{The physical picture of collinear photons emitted by classical source, $W_j$. The picture on the left is in the frame of the hard scattering and that on the right is in the frame where each component of the $j$-collinear momenta are the same size, $\lambda$.}
\label{fig:CollPhotSource}
	\end{center}
\end{figure}

We can make this story even closer to that of the soft by boosting in the $j$-th direction with a gamma factor of $\lambda^{-1}$. Then the $j$-cone becomes the whole space except a cone in the opposite direction which contains all of the radiation that was originally outside of the $j$-cone. Call this new cone the $\bar j$-cone. Now, all of the charges that were originally outside of the $j$-cone are in the $\bar j$-cone and act as a single classical source for radiation into the $j$-sector (which now fills almost the whole space, just as the soft sector did). The freedom of choice of the direction of the Wilson line, $t_j$, comes from the fact that any direction outside of the $j$-cone gets boosted into the $\bar j$-cone. This picture is shown in Figure~\ref{fig:CollPhotSource}.

\section{Factorization for general $S$-matrix elements}
\label{sec:IdentPart}
So far we have considered factorization for processes which can be written at leading-order as matrix elements of local hard-scattering operators $\cO(x)$ composed of $N$ fields. Factorization in the form on the right-hand-side of \Eq{genfactsQEDbox} actually holds more generally, for any process involving $N$ distinct directions, whether or not we represent the hard scattering as the matrix element of a local operator with $N$ fields. The generalization is perhaps easiest to see through an example.

Consider the scattering $\gamma \gamma \to \phi \phi^\star $ in scalar QED. At tree-level, there are 3-diagrams giving
\begin{align}
i \cM &= 
\quad
\fd{2cm}{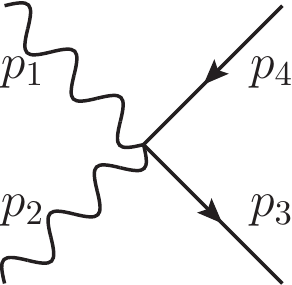}
 \quad+\quad 
\fd{2cm}{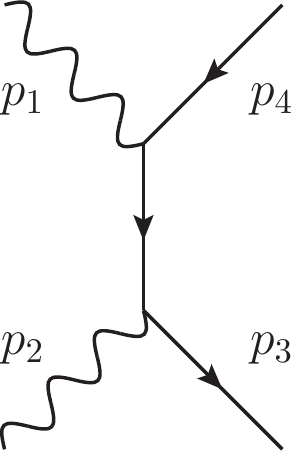}
 \quad+\quad 
\fd{2cm}{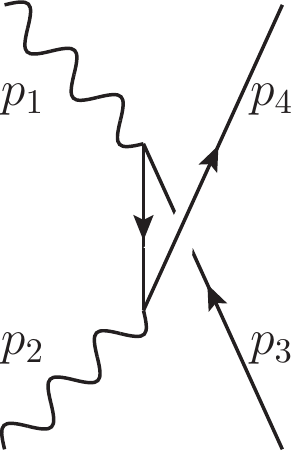} \\
& = 2i e^2\left[ (\epsilon_1 \cdot \epsilon_2)
 - \frac{(p_4 \cdot \epsilon_1)(p_3 \cdot \epsilon_2)}{p_4\cdot p_1}
 - \frac{(p_3 \cdot \epsilon_1)(p_4 \cdot \epsilon_2)}{p_3\cdot p_1}
\right]
\end{align}
There is not an easy way to write this amplitude as the matrix element of a gauge-invariant local operator $\cO$ with two scalar and two photon fields. The difficulty is that because the photons are identical, operators like
 $\frac{1}{p_1\cdot p_3} p_3^\mu p_4^\nu D_\mu \phi D_\nu \phi^\star$ give contractions with both photons. 
Factorization nevertheless holds for this process. In fact, as we will see, factorization holds for each diagram separately, as if it represents an independent hard process.

To begin, consider the most enhanced graphs when soft photons or photons collinear to $p_3$ or $p_4$ are added to the $t$-channel diagram. In generic-$r$, the only way to get a collinear enhancement is to have self-collinear emissions. Then, the only way to get a soft enhancement is by emission off of the 3 or 4 line because they are almost on shell. Thus, at leading power the $t$-channel part of the matrix element is given by a sum of graphs of the form
\be
\braket{X_3X_4;X_s | \epsilon_1(p_1)\epsilon_2(p_2)} \Big|_{t\text{-channel}} \;\LPeq\; \sum\quad \fd{3cm}{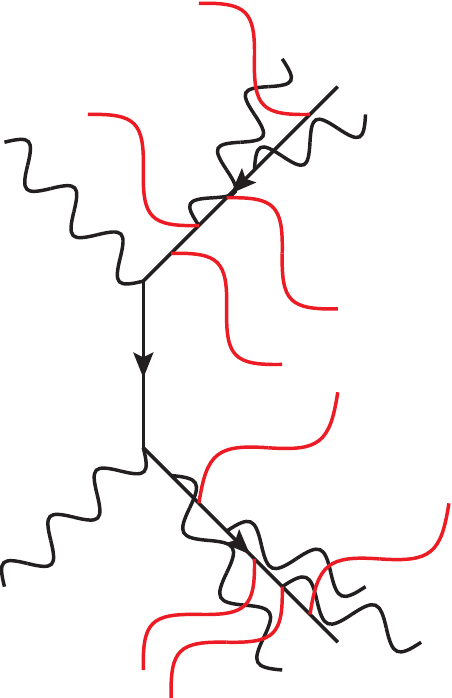}
\ee
Using the key soft-collinear factorization equation, \Eq{softpermuted}, we can strip the soft photons off of these graphs giving a factor of $\bra{X_s} Y_3^\dg  Y_4 \ket{0}$ multiplying the same graphs with no soft photons. Then, we use reference-vector
independence and collinear factorization, as in \Eq{collfactgen}, to write the collinear sectors as matrix elements of Wilson lines
$\bra{X_3} \phi^\star W_3  \ket{0}$ and $\bra{X_4} W_4^\dg \phi \ket{0}$.
The $u$-channel and 4-point diagram factorize in the same way.
We thus have 
\begin{multline}
\langle X_3 X_4 ; X_s | \epsilon_1(p_1)\epsilon_2(p_2)\rangle
\LPeq 
-2 i \left[ g^{\mu\nu}
 - \frac{P_4^\mu P_3^\nu}{P_4\cdot P_1}
 - \frac{P_3^\mu P_4^\nu}{P_3\cdot P_1}
\right]
\\
\times
\bra{0} \bar W_1^\dg D_\mu \bar{W}_1\ket{\epsilon_1(p_1)}
\bra{0} \bar W_2^\dg D_\nu \bar{W}_2\ket{\epsilon_2(p_2)}
\\
\times
 \bra{X_3} \phi^\star W_3 \ket{0} \bra{X_4} W_4^\dg \phi \ket{0} \bra{X_s} Y_3^\dg Y_4 \ket{0}
\end{multline}
where $P_3$ and $P_4$ are the total momenta of the states $\bra{X_3}$ and $\bra{X_4}$.
In  QED, we could have written $i e A_\mu$ instead of  $\bar W^\dg D_\mu \bar W$ but we write the matrix element this way so that the matching coefficient
only has dependence on momenta, independent of the spins. In QCD, similar factorized forms will arise with
$\bar W^\dg D_\mu \bar W$ reproducing matrix elements with multiple collinear partons in a gluon jet.

The result is that the $S$-matrix elements for this scattering process in scalar QED factorize. The factorization worked simply because soft and collinear emissions cannot couple to off-shell particles in Feynman diagrams at leading power (in generic-$r$). The same arguments apply to other scattering processes and to more complicated gauge theories like QCD. Thus, factorization holds for any hard scattering process with independent collinear sectors, irrespective of whether or not that process is written as the matrix element of a local operator.


\section{Spinor QED}
\label{sec:QED}
In the previous section,  soft-collinear factorization was proven (at tree-level)  in scalar QED. We now discuss how things change with spinors instead of scalars, and in the next section, go from QED to QCD.

Consider the following gauge-invariant hard-scattering operator in QED with $N$ flavors:
\be
\cO = \bar\psi_1\cdots \psi_N
\ee
We introduce the flavor indices on the fields only to simplify the contractions of the fields with states -- one can easily drop the subscripts.
We are interested in factorizing matrix elements of this operator in states comprising collinear and soft momenta, namely
\be
\bra{X_1\cdots X_N; X_s} \cO \ket{0}
\ee
Here we assume the flavor of the $i$-th jet matches the $i$-th field, for simplicity.
We continue to ignore collinear sectors initiated by a photon, which means that we do not consider covariant derivatives in the hard-scattering operator and the states $\bra{X_j}$ have 
the flavor quantum numbers of a single fermion.
Operators with derivatives or $\gamma$-matrices in the operator can easily be added, with the following proof hardly changing.
Collinear sectors initiated by a gauge boson will be treated in the QCD section where they are more interesting.

In this section we will show that the addition of spin does not affect the results found in scalar QED.
It does, however, require a little more notation.
 At leading order, when $\bra{X}= \bra{p_1\cdots p_N}$, the matrix
element is not 1 but rather
\be
\bra{p_1\cdots p_N} \cO \ket{0} = \bar{u}_1(p_1) \cdots v_N(p_N) \label{Ome}
\ee
where $u_i(p_i)$ or $v_i(p_i)$ are the particle or antiparticle spinor states contracted according to the fields
in $\cO$. A useful shorthand will be to pool everything that each spinor is contracted with into one object we denote $H^j$. Thus,
\be
\bra{p_1\cdots p_N} \cO \ket{0} = \bar{u}_1 H^1 = \cdots = H^N v_N 
\label{Hnotation}
\ee
That is, $H^j$ is just the hard-scattering matrix element with the $j$-th spinor stripped off.
In spinor-helicity notation
\be
\bra{p_1\cdots p_N} \cO \ket{0} = [p_1 H^1] = \langle p_2 H^2\rangle \cdots = [ H^N p_N]
\ee
with the bracket type depending on the helicity of the spinors, not whether it is particle or antiparticle.
Here we have used the freedom of little group scaling to choose the spinor helicity for the momentum $p_j$ to be exactly the spinor in the state $\bra{p_j}$.

\subsection{Collinear Factorization}
We start by considering only collinear photons; extra collinear spinors will come from insertions of the Lagrangian as in scalar QED and soft photons will be treated below. The derivation of collinear factorization is essentially the same as for scalar QED because in generic-$r$ exactly the same diagrams are enhanced. To see this, note that for a single collinear emission off of a massless spinor in QED, the leading order matrix element becomes
\begin{align}
\fd{3.5cm}{paperfigs/collsQED_vert.pdf} &= \frac{-g\, \bar u_j \Sl{\epsilon}_q(\Sl{p}_j+\Sl{q})}{2p_j\cdot q}\, H^j
= \frac{-g\, p_j\cdot\epsilon_q}{p_j\cdot q}\,  \bar u_j H^j
- \frac{g\, \bar u_j\Sl{\epsilon}_q\Sl{q}}{2p_j\cdot q}\, H^j
\label{twoterms}
\end{align}
where the Dirac equation $\bar{u}_j\, \slashed{p_j} = 0$ has been used. The first term is very similar to the scalar-QED result and follows the same story as the before: in generic-$r$ it only contributes at leading power
 when $q$ is collinear to $p_j$, whereas for the collinear-$r$ choice ($r=p_j$ for all photons collinear to $p_j$), the polarization vector itself is enhanced as in \Eq{refvectenhanced} and all \emph{but} the self-collinear emissions contribute.

The final term in \Eq{twoterms} is new. Since all the other (old) terms satisfy the Ward identity,
this term must satisfy the Ward identity by itself, which is easy to check:
\be
\frac{g \bar u(p_j)\Sl{\epsilon}_q\Sl{q}}{2p_j\cdot q}\,H^j \overset{\epsilon_q \to q}{=} 
\frac{g \bar u(p_j)\Sl{q}\Sl{q}}{2p_j\cdot q}\,H^j = \frac{g \bar u(p_j) q^2}{2p_j\cdot q}\,H^j = 0
\ee
Thus this term, by itself,  is reference-vector independent, which is also easy to check with
helicity spinors:
\be
\frac{g \bar u(p_j)\Sl{\epsilon}_q\Sl{q}}{2p_j\cdot q}H^j
= \sqrt{2}g\frac{[p_j q]\l r q\r [q H^j] }{\l qr\r[p_jq]\l qp_j\r}
= \sqrt{2}g\frac{[q H^j] }{\l p_jq\r}
\ee
So, independently of the reference-vector choice, the new term will only contribute at leading power when $q \parallel p_j$, that is, for self-collinear emissions. Such emissions will come from a field emitting
a photon through a Lagrangian interaction (as opposed to from a Wilson line), just as they do in the unfactorized expression.

Thus, we have exactly the same diagrammatic factorization as in \Eq{diagfactsQED} and by the same gauge-symmetry arguments, we get the same result as \Eq{collfactgen}, namely
\be
\bra{X_1\cdots X_m} \bar\psi_1\cdots \psi_m \ket{0}
\LPeq
\bra{X_1}\bar\psi_1 W_1 \ket{0}
\dots
\bra{X_m}W_m^\dg\psi_m \ket{0}
\label{collfactQED}
\ee 
 The right-hand side of this equation reproduces \Eq{twoterms} for one emission.
 More generally, terms like  $\frac{\Sl{\epsilon}_q\Sl{q}}{p_j\cdot q}$ will always come from 
 Lagrangian emissions, while the the eikonal terms, $\frac{p_j\cdot \epsilon_q}{p_j\cdot q}$, either
 come from the Lagrangian (in generic-$r$) or the Wilson lines (in collinear-$r$).

\subsection{Soft and soft-collinear factorization}
\label{sec:QEDSF}

The addition of spin has no affect on the factorization of soft emissions because soft emissions cannot flip the spin of non-soft particles. For example, in the soft limit of QED
\be \label{softQEDvertex}
\fd{3.5cm}{paperfigs/softsQED_vert.pdf} = \frac{-g \bar u_j\Sl{\epsilon}_k(\Sl{p}_j+\Sl{k})}{2p_j\cdot k }
H^j
\LPeq -g\,\frac{ p_j\cdot\epsilon_k}{p_j\cdot k }\, \bar u_j H^j
\ee
This vertex is identical to the scalar QED result in \Eq{sQEDsoft1}. 
Since the soft limit is spin independent, soft factorization is identical in QED and scalar QED.

More generally, a useful fact is that the soft limit of the matrix element for a photon interacting with a particle of any spin or mass has the same eikonal form. The physical reason is simply that an arbitrarily soft photon does not have enough energy to flip the helicity of a particle. 
A proof proceeds as follows: let $\zeta_s^\alpha(p)$ be the wavefunction for a particle of mass $m$ and helicity $s$ that interacts with a gauge boson through a current $J^\mu$. For example, for spin 1, $\zeta^\alpha_s(p) = \epsilon_h^\mu(p)$ are the polarization vectors and for spin $\frac12$, $\zeta^\alpha_s(p) = u_s^i(p)$ are the Dirac spinors. By unitarity, we can always write the two-point function for this particle as a sum of dyads:
\be
G^{\alpha\beta}(x,y) = \int d^4p \, e^{i p (x-y)} \,
	\frac{ i\sum_s \zeta_s^\alpha(p) \zeta_s^\beta(p)^\dagger}{p^2 - m^2 + i \varepsilon}
\label{dyads}
\ee
The key ingredient for the proof of spin independence is that the vertex for the emission of a soft gauge boson from an on-shell $\zeta$ particle is of the form~\cite{Weinberg:1964ew}:
\be
\bra{p, s'} J^\mu(0) \ket{p, s}
= \zeta_s^\alpha(p)^\dg \;\Gamma^\mu_{\alpha\beta}\, \zeta_{s^\prime}^\beta(p)
= 2\, p^\mu\, \delta_{ss'}
\label{univsoft}
\ee
The last equality is the statement of helicity conservation. 
Combining Eqs.~\eqref{dyads} and \eqref{univsoft}, we get
\begin{align}
\fd{3.5cm}{paperfigs/softsQED_vert.pdf} \;&\LPeq  ig\,\epsilon_{k\,\mu}\,\bar u_s(p_j) \cdot\Gamma^\mu \cdot \frac{i\sum_{s'} u_{s'}(p_j)\bar u_{s'}(p_j) }{(p_j+k)^2 - m^2} H^j  \notag
\\&=  ig\,\epsilon_{k\,\mu}\, 2 p_j^\mu\,  \frac{i\sum_{s'} \delta_{ss'}\bar u_{s'}(p_j) }{2p_j\cdot k} H^j 
	\label{univsoftuse}
\\& =  -g\, \frac{ p_j\cdot\epsilon_{k}}{p_j\cdot k} \,\bar u_{s}(p_j) H^j \notag
\end{align}
Thus, the eikonal form of the soft interaction holds for any mass and spin.
For non-Abelian gauge bosons this expression just gets multiplied by a generator matrix.
Although this proof may seem pedantic for QED where the eikonal form can be derived much more directly, the spin-independence of the soft limit is very useful more generally. For example, it is actually quite cumbersome to show the eikonal form for a soft gluon emitted from a collinear gluon in QCD. Thus, \Eq{univsoft} will be put to pragmatic use in the next section.

Since soft factorization is identical in spinor QED as in scalar QED, the derivation of soft-collinear factorization
from Section~\ref{sec:sQEDSSCF} goes through unchanged. 
Therefore, in QED the same soft-collinear factorization formula as
in \Eq{genfactsQEDbox} holds, with the replacement $\phi \to \psi$:
\be
\begin{array}{|c|}
\hline\\
~~~\bra{X_1\cdots X_N; X_s} \cO \ket{0} 
\LPeq \bra{X_1}\bar \psi_1 W_1 \ket{0}
\dots \bra{X_N}W_N^\dg\psi_N \ket{0}\bra{X_s} Y_1^\dg\cdots Y_N \ket{0}
~~~\\
\\
\hline
\end{array}
\ee


\section{QCD}
\label{sec:QCD}
We are now ready to tackle the final details relevant for factorization in QCD. We wish to factorize matrix elements of a gauge-invariant hard-scattering operator of the form
\be
\cO^\mu = \bar\psi_1\cdots \bar\psi_{m-1} (D^\mu) \psi_{m+1} \cdots \psi_{N}
\ee
with $N-1$ spinors and one covariant derivative. 
Here, as in the spinor QED section, the subscripts on the spinor fields are flavor indices added only to simplify the combinatorics. Color
indices are suppressed to avoid clutter (cf. \Eq{withcolor} below).
Removing the flavor indices or considering more than a single covariant derivative requires us to keep track of tedious combinatoric factors and contractions among different spinors which dirty the expressions in our factorization proof but do not change the results in any substantial way.
Such cases are best dealt with in a similar fashion to that described in Section~\ref{sec:IdentPart}.
The covariant derivative $D^\mu$ in the operator is an easy way to give the operator non-zero 
matrix elements in a state with a gluon in a particular direction. We consider matrix elements of this operator between the vacuum and the $N$-jet final state $\bra{X} = \bra{X_1\cdots X_N;X_s}$,
with each state $\bra{X_j}$ either a quark jet, with the flavor of a single quark species, or the gluon jet which we place in $\bra{X_m}$. 
The matrix element of this operator in the simplest such state, with $N-1$ spinors and one gluon is,
\begin{align}
\bra{p_1\cdots p_m,a\cdots p_N} \cO^\mu \ket{0} 
&= \bar{u}_1(p_1) \cdots 
		(-igT^a)\epsilon^\mu(p_m)
		\cdots v_N(p_N) \\
&\equiv \bar{u}_1(p_1)H^\mu_1 = \cdots = \epsilon^\mu(p_m) H_m
\end{align}
as in \Eq{Ome} or \Eq{Hnotation}.
We are considering the simplest possible operator, $\cO$, for clarity. Operators with nontrivial Dirac structure and insertions of partial derivatives change nothing but the form of the above $H_i$'s.

\subsection{Collinear Factorization}
\label{sec:QCDCF}

For gluons emitted off of a quark line, collinear factorization follows immediately from the generic-$r$ choice of reference vectors. In
generic-$r$ only self-collinear emissions are enhanced.  That is, 
\be
\fd{3.5cm}{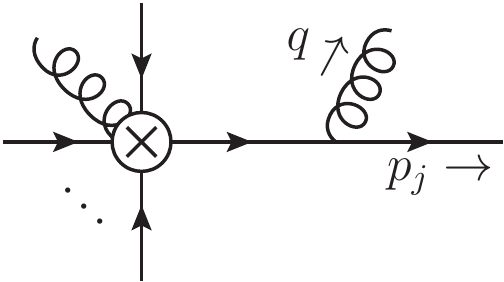} 
 \overset{\text{gen.}~r}{\sim}
 \bar{u}_j (g \lambda^{-1}) H^\mu_j 
\ee
while
\be
\fd{3.5cm}{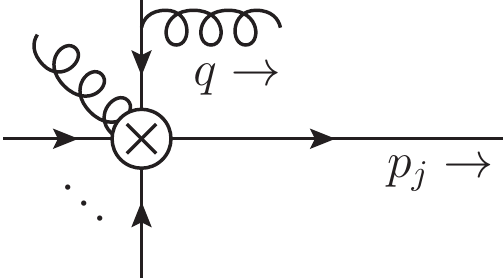}  
 \overset{\text{gen.}~r}{\sim}
\bar{u}_j( g \lambda^0)H^\mu_j 
\ee
Although the momentum going into the hard vertex depends on $q+p_j$, this induces no additional enhancement. 

For a gluon jet, collinear emissions follow the same pattern. In generic-$r$, self-collinear emissions
\be
\begin{array}{ccc}
\fd{3.5cm}{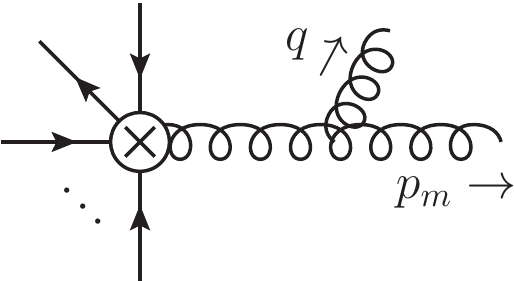}
 \overset{\text{gen.}~r}{\sim}
\dfrac{g}{\lambda} \epsilon^\mu H_m
&
\quad\text{and}\quad
&
\fd{3.5cm}{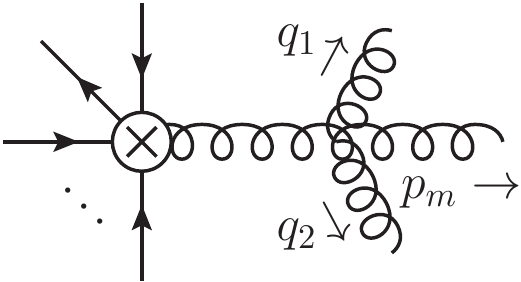}
 \overset{\text{gen.}~r}{\sim}
\left(\dfrac{g}{\lambda}\right)^2 \epsilon^\mu H_m
\label{collcountQCD}
\end{array}
\ee
dominate over emissions off of other legs
\be
\fd{3.5cm}{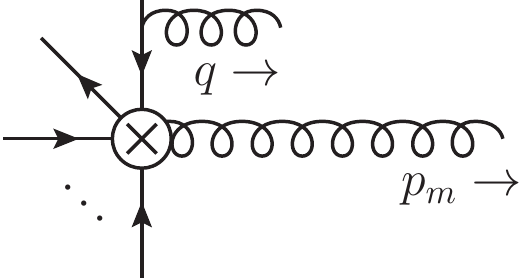} 
 \overset{\text{gen.}~r}{\sim}
g \lambda^0 \epsilon^\mu H_m
\ee
Thus, only self-collinear emissions are relevant at leading power, for either quark or gluon jets.

Collinear factorization in generic-$r$ is therefore identical in QCD and QED. We can write the result as
 \be
\bra{X_1\cdots X_N} \cO^\mu \ket{0}
 \overset{\text{gen.}~r}{\LPeq}
 \bra{X_1}\bar \psi_1^{l_1} \ket{0} \cdots \bra{X_m} (D^\mu)^{l_{m-1} l_{m+1}} \ket{0}
\dots \bra{X_N} \psi_N^{l_N} \ket{0} 
\label{withcolor}
\ee
Here, we have displayed the fundamental color indices $l_1 \cdots l_N$ explicitly.
These indices are all contracted since the original operator $\cO$ was gauge invariant.
The notation $(D^\mu)_{i j} \equiv \delta_{ij} \partial^\mu - i g T^a_{ij} A^{a\mu}$ is the usual covariant derivative
in the fundamental representation.

The reference-vector independent leading-power result is the generalization of \Eq{collphotfact}:
\begin{multline}
\bra{X_1\cdots X_N} \cO^\mu \ket{0} \LPeq \\
\bra{X_1}\bar \psi_1^{l_1} W_1^{l_1 h_1} \ket{0} \cdots 
\bra{X_m} W_m^{\dg\, h_{m-1} l_{m-1}} (D^\mu)^{l_{m-1} l_{m+1}} W_m^{l_{m+1} h_{m+1}}\ket{0}
\dots \bra{X_N} W_N^{\dg\, h_N l_N}\psi_N^{l_N} \ket{0} 
\label{QCDcollfact}
\end{multline}
where now the $W$ are non-Abelian path-ordered Wilson lines and the new $h_i$ color indices are summed over. 
One might be concerned that because the product $W^{\dg\, h l}\psi^{l}$ carries a color index, $h$, it is not gauge invariant. However, the $h$ index
affects the transformation properties at $x=\infty$ which are trivial.

Finally, to make contact with our expectations from the physical intuition that a gluon jet should see a collinear Wilson line in the adjoint representation coming from the rest of the hard scattering process, we 
can use \Eq{YYrel}, $ W_n^\dg A^\mu_b T^b\, W_n = A^\mu_a \,\cW_n^{ab}\, T^b$.
So, it is indeed the case that the radiation from the rest of the event into a jet initiated by a gluon, $A_a$, in the collinear limit appears as if coming from a classical source of the form of a Wilson line in the adjoint representation, $\cW^{ab}$.

\subsection{Soft Factorization}
 \label{sec:QCDSF}

The factorization of soft gluons off of the hard scattering matrix element is only different from the scalar-QED case in that we cannot use the eikonal identity as in \Eq{softemisqline} because the generator matrices do not commute. However, this changes nothing since the Wilson line is path ordered and exactly makes up for this:
\begin{align}
& \sum_\text{perms}\;\fd{7.5cm}{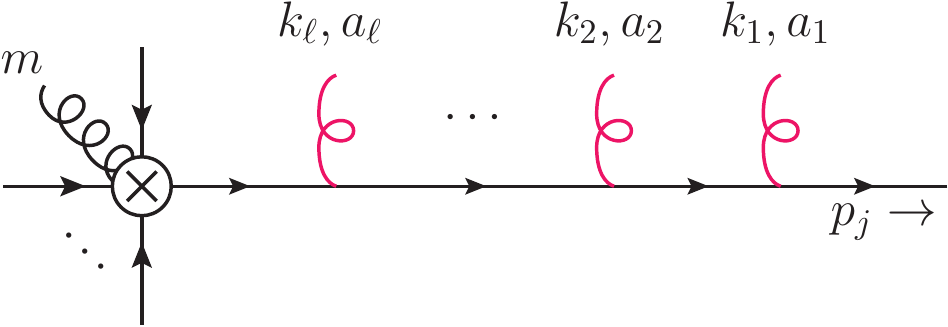} \notag
\\ &\hspace{1cm} = \bar u(p_j) \bigg(\sum_\text{perms}(-g)^\ell \, \frac{n_j\cdot \epsilon_1}{n_j\cdot k_1}T^{a_1} \frac{n_j\cdot \epsilon_2}{n_j\cdot (k_1+k_2)} T^{a_2} \cdots \frac{n_j\cdot \epsilon_\ell}{n_j\cdot \sum_{i=1}^\ell k_i} T^{a_\ell} \bigg)H^\mu_j
	\label{softYQCD1}
\\ &\hspace{1cm} =\bar u(p_j)  \bra{k_1\cdots k_\ell} Y_j^\dg(0) \ket{0} H^\mu_j \notag
\end{align}
Similarly, we can use general soft gauge-boson vertex of \Eq{univsoft} as was done in \Eq{univsoftuse} to show that
\begin{align}
& \sum_\text{perms}\;\fd{7.5cm}{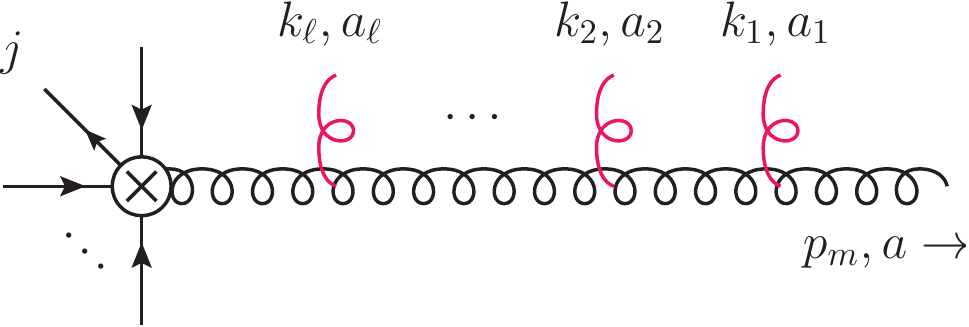}  
	\notag
\\ &\hspace{1cm} = \bigg(\sum_\text{perms}(-g)^\ell \, \frac{n_m\cdot \epsilon_1}{n_m\cdot k_1}T_\text{adj}^{a_1}\, \frac{n_m\cdot \epsilon_2}{n_m\cdot (k_1+k_2)} T_\text{adj}^{a_2}\, \cdots \frac{n_m\cdot \epsilon_\ell}{n_m\cdot \sum_{i=1}^\ell k_i} T_\text{adj}^{a_\ell} \bigg)\; \epsilon^{\mu}_{p_m} H_m
	\label{softYQCD2}
\\ &\hspace{1cm} = \bra{k_1\cdots k_\ell} (\cY_m^\dg(0))^{ab} \ket{0}\;
	\epsilon^{\mu}_{p_m} H^b_m 		\notag
\end{align}
where $\cY_i$ is the usual soft Wilson but in the adjoint representation and in the last line we wrote the adjoint color indices explicitly.

\Eq{softYQCD1} and \eqref{softYQCD2} are the QCD equivalents of \Eq{keyeqSF}, which as mentioned in Section~\ref{sec:sQEDSF} is all we need to show, since all of the arguments in that section were completely general. Thus, using the arguments of Section~\ref{sec:sQEDSF}, we arrive at the QCD equivalent of \Eq{softfactsQED}, namely
\begin{multline}
\bra{p_1\cdots p_N;k_1\cdots k_\ell} \bar\psi_1^{l_1} \cdots (D^\mu)^{l_{m-1}l_{m+1}} \cdots \psi_N^{l_N} \ket{0}
\\=
\bra{p_1\cdots p_N} \bar\psi_1^{l_1} \cdots (D^\mu)^{l_{m-1}l_{m+1}} \cdots \psi_N^{l_N} \ket{0}
\\\times
\bra{k_1\cdots k_\ell} Y_1^{\dg\,l_1h_1} \cdots Y_m^{h_{m-1}l_{m-1}} Y_m^{\dg\,l_{m+1}h_{m+1}} \cdots Y_N^{h_Nl_N} \ket{0}
\label{QCDsoftfact}
\end{multline}
Note that one does not have to go through all of the arguments of Section~\ref{sec:sQEDSF} to figure out the contractions of indices in the equation above. For example, \Eq{softYQCD1} shows that a collinear quark field, $\bar\psi^h$ should become $\bar\psi^l Y^{\dg lh}$. Also, one does not need to worry about how the derivative in $D^\mu$ acts on the soft Wilson line because it is power suppressed.

We can now do the usual replacement $\bra{k_1\cdots k_\ell} \to \bra{X_s}$, allowing $\bra{X_s}$ to contain soft quarks, because any soft quarks must come soft-gluon splitting at leading power.

\subsection{Soft-Collinear Factorization}
\label{sec:QCDSCF}

The simultaneous soft and collinear factorization in QCD parallels that of scalar QED completely; the only difference being the more complicated non-Abelian charges of QCD. For example, the soft coherence of \Eq{softcoher} is the same. Imagine taking $k\to 0$ naively as in \Eq{softcoher} where we ignore $k$ in any internal lines. Then taking $\cM = 1$, $P = p+q_1+q_2$ and the collinear-gluon three-point vertex to be $-iV_3^{\rho\alpha\beta}f^{ebc}$, we have:
\begin{align}
&\fd{5cm}{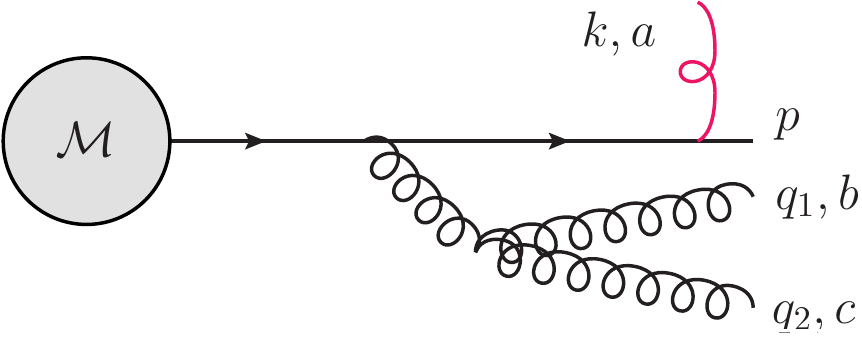} \,+\, \fd{5cm}{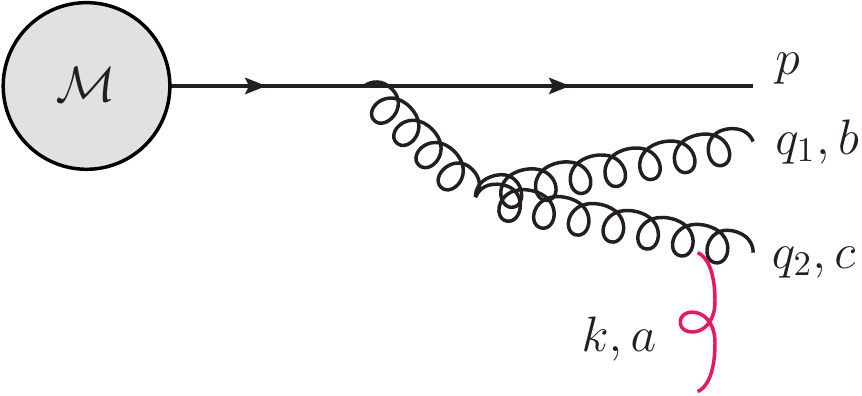} \;+\; \fd{5cm}{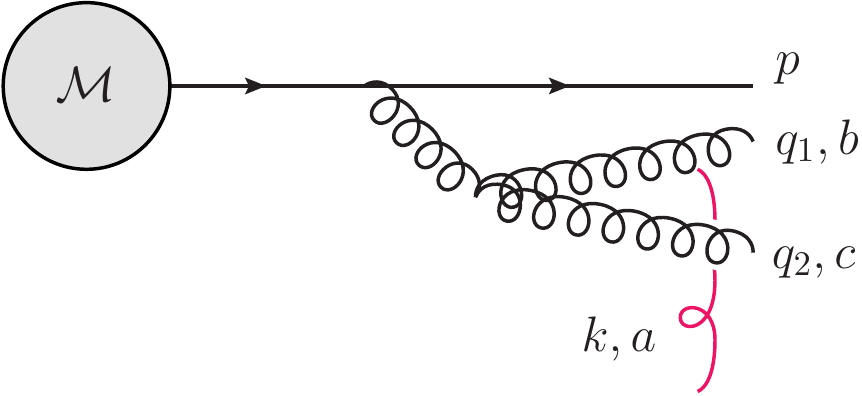}
\notag
\\&= \bar u_{p}  \frac{g\gamma_\rho}{(q_1+q_2)^2} \frac{\Sl{P}}{P^2}
\,V_3^{\rho\alpha\beta}\, 
\bigg[ 	t^at^e f^{ebc}\,\frac{-g\, p\cdot\epsilon_k }{p\cdot k} 
	   -	t^ef^{ebc'} \frac{g\, q_2\cdot\epsilon_k (T_\text{adj}^a)_{cc'} }{q_2\cdot k} 
	   -	t^ef^{eb'c} \frac{g\, q_1\cdot\epsilon_k (T_\text{adj}^a)_{bb'} }{q_1\cdot k}	\bigg] \notag
\\&= \bar u_{p}  \frac{g\gamma_\rho}{(q_1+q_2)^2} \frac{\Sl{P}}{P^2}
\,V_3^{\rho\alpha\beta}\, 
\Big[ t^et^a f^{ebc} + it^d \big( f^{aed} f^{bce} + f^{bed}f^{cae} + f^{ced}f^{abe} \big)\Big] 
	\frac{-g\,n\cdot \epsilon_k}{n\cdot k} \notag
\\&= \bar u_{p}  \frac{g\gamma_\rho\,t^e}{(q_1+q_2)^2} \frac{i\Sl{P}}{P^2}
\,\big(-iV_3^{\rho\alpha\beta}f^{ebc} \big)\, \frac{-g\,n\cdot \epsilon_k\,t^a}{n\cdot k}
\label{softcoherQCD}
\end{align}
To get the second line we used the general soft vertex of Section~\ref{sec:QEDSF} and to get the last equation we used the Jacobi identity. The last line shows that the soft gluon only sees the total charge of the collinear sector.

However, as in \Eq{softcoher}, this only works if we assume the soft gluon momentum $k$ is much softer than all the other momenta, which is too strong of a restriction.  In fact $k$, can be of the same order as the $p\cdot q_i \sim \cO(\lambda^2)$. Hence, we again have to  worry about the tangled diagrams, which means we must also consider:
\be
\fd{4.4cm}{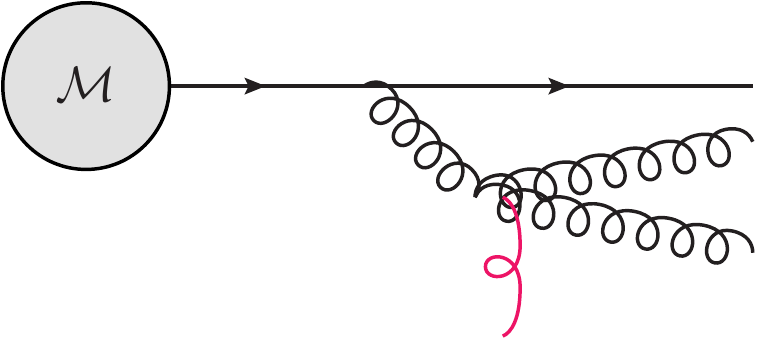} \,+\, \fd{4.4cm}{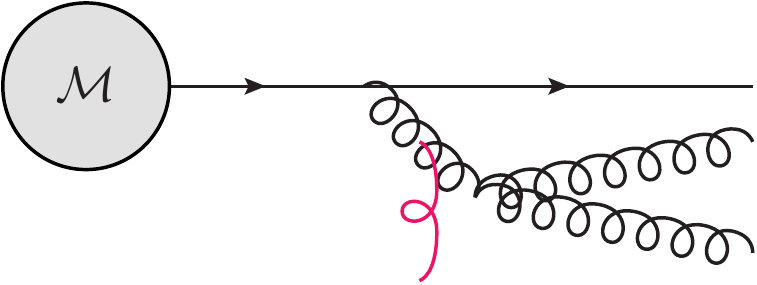} \;+\; \fd{4.4cm}{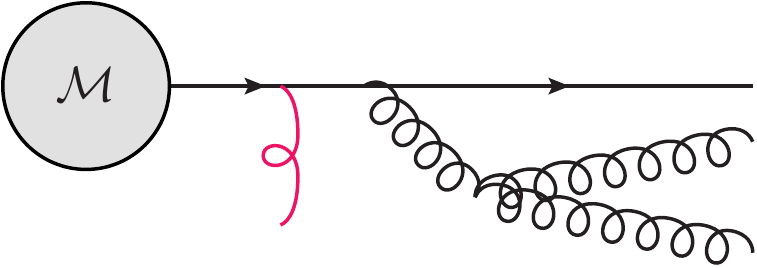}
\ee
Indeed, summing these three diagrams with those in \Eq{softcoherQCD} would give the eikonal form for the soft emission, but we want to prove soft-collinear factorization more generally.

Luckily we have already done so! Section~\ref{sec:sQEDGSCF} used nothing about scalar QED; it only used that soft factorization and collinear factorization had been shown on their own. We have already shown both collinear and soft factorization separately in sections \ref{sec:QCDCF} and \ref{sec:QCDSF}, so all we need to do is go through Section~\ref{sec:sQEDGSCF} step by step to get a general proof for QCD.

The only extra detail of QCD is the color indices which we already know how to contract from \Eq{QCDsoftfact}. Therefore, we can simply write down \Eq{gensoftfact} for QCD as:
\begin{multline}
\bra{X_1\cdots X_N;X_s} \bar\psi_1^{l_1} \cdots (D^\mu)^{l_{m-1}l_{m+1}} \cdots \psi_N^{l_N} \ket{0}
\\=
\bra{X_1\cdots X_N} \bar\psi_1^{l_1} \cdots (D^\mu)^{l_{m-1}l_{m+1}} \cdots \psi_N^{l_N} \ket{0}
\\\times
\bra{X_s} Y_1^{\dg\,l_1h_1} \cdots Y_m^{h_{m-1}l_{m-1}} Y_m^{\dg\,l_{m+1}h_{m+1}} \cdots Y_N^{h_Nl_N} \ket{0}
\label{QCDgensoftfact}
\end{multline}
Now that the soft gluons are factorized, we use Section~\ref{sec:QCDCF} (generic-$r$ and reference-vector independence), in particular \Eq{QCDcollfact}, to get the general soft-collinear factorization in QCD:
\begin{multline}
\begin{array}{|c|}
\hline\\
\hspace{-6.5cm}\bra{X_1\cdots X_N;X_s} \bar\psi_1^{l_1} \cdots (D^\mu)^{l_{m-1}l_{m+1}} \cdots \psi_N^{l_N} \ket{0}
\\[2mm] \hspace{-.5cm}=
\bra{X_1} (\bar \psi_1W_1)^{l_1} \ket{0} \cdots 
\bra{X_m} (W_m^{\dg} D^\mu W_m)^{l_{m-1} l_{m+1}}\ket{0}
\dots \bra{X_N} (W_N^{\dg}\psi_N)^{l_N} \ket{0} 
\\[2mm] \hspace{5.5cm} \times
\bra{X_s} Y_1^{\dg\,l_1h_1} \cdots Y_m^{h_{m-1}l_{m-1}} Y_m^{\dg\,l_{m+1}h_{m+1}} \cdots Y_N^{h_Nl_N} \ket{0}
\\ \\
\hline
\end{array}
\label{genfactQCD}
\end{multline}
Note that the $l_j$ index on each of the collinear matrix elements and on each soft Wilson line transforms at infinity, so each term in this equation is separately gauge invariant.

\Eq{genfactQCD} is our final result and has been proven at tree-level for arbitrary collinear states $\bra{X_j}$ either initiated by a quark or a gluon and an arbitrary soft state $\bra{X_s}$. It is the statement that, when considering the scattering of energetic massless particles interacting with gauge bosons, the form of soft and collinear emissions simplifies tremendously. Hard-scattered particles see only a collinear Wilson line with charge opposite their own in place of all possible collinear gauge-boson emissions from the rest of the scattering. Furthermore, they emit soft gauge bosons in the form of a classical source moving in their direction of travel  with their charge.


\section{SCET}
\label{sec:SCET}

To touch base with SCET, in particular the formulation in~\cite{Freedman:2011kj}, we note that each
matrix element on the right-hand side of \Eq{genfactQCD} can be computed with a separate
copy of the QCD Lagrangian. We can formalize this by writing an effective Lagrangian
which is the sum of $N+1$ copies of the QCD Lagrangian:
\be
\cL_\text{eff} = \cL_\text{soft} + \sum_{j=1}^N \cL_j
\label{newSCETLag}
\ee
Then we assign separate quantum numbers $j$ or $s$ to the particles in $X_j$ and $X_s$ associated with their sector, with the fields in the Lagrangians $\cL_j$ or $\cL_{\text{soft}}$ being only able to create
or destroy particles with the appropriate quantum number. Once this is done, we can simply combine all
the matrix elements together to write
\begin{multline}
\bra{X_1\cdots X_m; X_s} \bar\psi_1\cdots D^\mu \cdots \psi_N \ket{0} \\
	=\bra{X_1\cdots X_m; X_s} \big(\bar\psi_1\, W_1\, Y_1^\dg\big) \cdots
\big( Y_m \, W_m^\dg \, D^\mu \, W_m \, Y_m^\dg \big)
\cdots \big(Y_N\, W_N^\dg\, \psi_N\big) \ket{0}_{\cL_\text{eff}} 
\label{FLSCET}
\end{multline}
In this way both sides are matrix elements of an operator with a Lagrangian. In this form,
the agreement can be pursued beyond tree-level with corrections absorbed into a finite
Wilson coefficient on the right-hand side.

This formulation of SCET is most similar to the Luke/Freedman formulation~\cite{Freedman:2011kj}, 
which partly inspired the current work, but even simpler
since we do not attempt to make the agreement palatable to the inclusion of power corrections. 
More explicitly, the Luke/Freedman formulation has fields in the operators evaluated at different positions, such as 
$x_{\bar{n}} \equiv (n\cdot x, 0,\vec{x}_\perp)$, as in the multipole formulation of SCET~\cite{Beneke:2002ph,Beneke:2002ni}.  The simple way to see why
soft and collinear fields interact only at $x_{\bar{n}}$ is that these positions are within the jet cone,
which is where the soft and collinear radiation can overlap. However, as long
as the Lagrangian is defined to contain decoupled sectors, changing the location at which the fields are evaluated in this way only affects subleading powers. 

Both the Luke/Freedman and the multipole formulation of SCET differ somewhat from label SCET~\cite{Bauer:2000ew,Bauer:2000yr,Bauer:2001yt}. In label SCET, fields are all evaluated at the same point, but the interactions are not those of full QCD. Instead there are an intricate set of SCET Feynman rules,
derived by integrating out large components of spinors, as is done in heavy quark effective theory,
and then removing certain interactions through field redefinitions.  For soft interactions, these rules
are the eikonal Feynman rules. For collinear sectors, the rules are equivalent to QCD in light-cone gauge~\cite{Beneke:2002ph}. It has already been observed that label SCET, after
a field redefinition which decouples the soft from the collinear interactions in the Lagrangian, is equivalent to having multiple copies of QCD~\cite{Bauer:2008qu}. 

We will not attempt to explain, justify or defend any formulation of SCET in this paper. As far as anyone can tell, all the formulations are equivalent at leading power. The point of this section is merely to reiterate the observation of~\cite{Freedman:2011kj} that factorization can be phrased in terms of QCD fields and a  Lagrangian which contains multiple independent copies of QCD. Indeed, the point of this paper
is essentially to give a transparent proof of the observations made in~\cite{Freedman:2011kj},
using on-shell methods.


\section{Application: the QCD Splitting Functions}
\label{sec:splitting}

As an application of the factorized expressions that we have derived, we will compute the tree-level splitting functions in QCD. Perhaps the simplest way to compute unpolarized splitting functions is following Altarelli and Parisi~\cite{Altarelli:1977zs} by squaring the relevant three-point vertex and summing over spins.
For this sum, they use the replacement
  \be
  \sum_{\text{pols}} \epsilon_q^{*\nu} \epsilon_q^\mu 
  \to
  \begin{cases}
  \delta^{ij} - \frac{q^i q^j}{q^2} & i,j >0
  \\ 0 & \text{otherwise}
  \end{cases}
  \label{APreplace}
\ee
It is important to use this replacement, and not the simpler $  \sum_{\text{pols}} \epsilon^{*\nu} \epsilon^\mu  \to -g^{\mu\nu}$, since the simpler replacement assumes the Ward identity is satisfied,
which is not the case for a single emission off of a single leg of a matrix element. The replacement
in \Eq{APreplace} is equivalent to a sum over polarization vectors with reference vector choice 
$r^\mu = (1,0,0,-1)$. Instead, if one chooses the reference vector to be in the direction of the quark that splits, 
one would find a different answer for the splitting functions: zero.

In the on-shell language we have been advocating, the correct cross section
to evaluate is   $|\bra{ p; q}\bar{\psi}\, W_t \ket{0}|^2$. In this case, summing over polarizations
will be independent of reference vector, even if $r \sim p$. Of course, the usual calculation using physical polarizations produces the right answer since in this case the contributions from the Wilson line
are power suppressed. It is nevertheless illustrative (and easy) to see the computation performed
using spinor helicity methods.
Moreover, since we have already shown that $|\bra{ p; q} \bar{\psi}\,W_t\ket{0}|^2$ factorizes
off from any matrix element when $q$ becomes collinear to $p$ this approach automatically also proves
the universality of splitting functions. The connection between the splitting functions and collinear factorization in effective field theory was also observed in~\cite{Bauer:2006mk,Bauer:2006qp}.

\subsection{Quark-gluon splitting function}
We have shown that for a generic process with  a collinear sector initiated by a quark, the matrix element factorizes into $\bra{X_j} \bar\psi\, W_j \ket{0} $  times a matrix element that has no collinear radiation in the $n_j$ direction. Here, $W_j$ points in some direction $t^\mu$ not collinear to $n_j$.
  Therefore, the quark-gluon splitting amplitude is given by:
\be
\bra{p;q,a} \bar\psi\, W_j \ket{0} \;=\quad \fd{2.3cm}{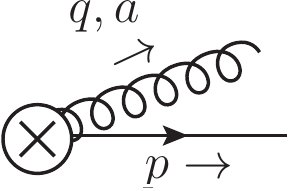} \quad+\quad \fd{2.9cm}{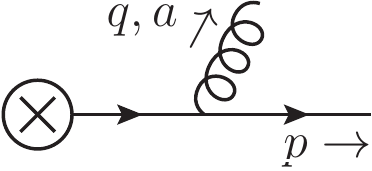}
\ee
times the rest of the amplitude, $\cM(p+q)$, which carries a color and spinor index. That is,
\be 
\cM_\text{split} = \bar u(p)\bigg[
\frac{t \cdot\epsilon(q)}{t \cdot q}
-
\frac{p\cdot\epsilon(q)}{p\cdot q}
-
\frac{\cn{\epsilon}(q)\cn{q}}{2p\cdot q} 
\bigg] g\,T^a\cM(P)
\ee
where $P = p+q$.
The nice thing about this expression is that, because it satisfies the Ward identity, we can choose any reference vector for $\epsilon^\mu(q)$. Moreover, since we have proven collinear factorization for any hard-scattering operator, we know that this expression is the universal collinear-splitting amplitude.

Instead of doing the calculation in generic-$r$, like in \Eq{APreplace},
we use collinear-$r$. Thus we take $r=p$. Then,
\be
\epsilon_q^- = \sqrt{2}\,\frac{p]\l q}{[qp]} 
\quad\And\quad
\epsilon_q^+ = \sqrt{2}\,\frac{q]\l p}{\l pq\r}
\ee
both of which satisfy
\be
p\cdot\epsilon(q)=0
\ee
Let us also take the spinor to be right-handed, namely $\bar u(p) = [p$. Then for $\epsilon^-$ we find
\be
\cM_\text{split}^{R,-} = \sqrt{2}\, g\frac{ [t p]}{[t q][qp]} [p\,T^a\cM 
= \sqrt{2}\, \frac{g}{[qp]} \frac{z}{\sqrt{1-z}} \,[P\,T^a\cM 
\ee
and for  $\epsilon^+$ we find 
\be
\cM_\text{split}^{R,+} 
= \sqrt{2}\, g\bigg( \frac{\l p t\r}{\l q t\r \l pq\r}[p - \frac{1}{\l qp\r }[q \bigg)T^a\cM
=  \sqrt{2}\, \frac{g}{\l pq\r}\bigg( \frac{z}{\sqrt{1-z}} + \sqrt{1-z} \bigg) [P\,T^a\cM
\ee
where we have used that
\be
p^\mu = z\, P^\mu \quad\And\quad q^\mu = (1-z)\, P^\mu
\ee
which implies that
\be
[p = \sqrt{z}\,[P, \qquad [q = \sqrt{1-z}\,[P, \qquad \text{ etc}
\ee
at leading power. 

Using $\cM_\text{split}$, we can write down the polarized splitting functions:
\be
\big|\cM^{R, -}_\text{split} \big|^2 = \frac{g^2 C_F}{p\cdot q} \frac{z^2}{1-z}\, \big|\cM_\text{orig}\big|^2
\And
\big|\cM^{R,+}_\text{split} \big|^2 = \frac{g^2 C_F}{p\cdot q} \frac{1}{1-z}\, \big|\cM_\text{orig}\big|^2
\ee
where $\big|\cM_\text{orig}\big|^2 = \big|[P \cM(P)\big|^2$ is the original amplitude without collinear splitting. By parity invariance, we must also  have
\be
\big|\cM^{L,-}_\text{split} \big|^2 = \frac{g^2 C_F}{p\cdot q} \frac{1}{1-z}\, \big|\cM_\text{orig}\big|^2
\And
\big|\cM^{L,+}_\text{split} \big|^2 = \frac{g^2 C_F}{p\cdot q} \frac{z^2}{1-z}\, \big|\cM_\text{orig}\big|^2
\ee
The unpolarized splitting function is given by the sum of the two:
 \be
 \big|\cM^\text{unpol}_\text{split} \big|^2 = \frac{g^2 C_F}{p\cdot q} \frac{1+z^2}{1-z}\, \big|\cM_\text{orig}\big|^2
 \ee
which is the familiar result.

\subsection{Gluon-gluon splitting function}
Next, consider the gluon splitting function. In this case, we want to relate the cross section for the emission of two on-shell gluons, of momenta $p^\mu$ and $q^\mu$ to the amplitude
 $\cM^{c,\mu}_\text{orig}(p+q)$
  for producing a single off-shell gluon of momentum $p^\mu+q^\mu$.
We have already shown that for the collinear emission of gluons off of gluons,
the amplitude in QCD is reproduced by the matrix element  $\bra{X_j}W_t^\dg D^\mu W_t\ket{0}$
at leading power. Thus, if two $n$-collinear gluons of color $a,b$, momenta $p,q$ and polarizations $\epsilon_p$ and $\epsilon_q$ are omitted, the original amplitude gets modified to
\begin{multline}
\cM^{ab}_\text{split} = -igf^{abc}\bigg[
\epsilon_p^{*\mu} \, \frac{t \cdot \epsilon_q^*}{t\cdot q}
- \epsilon_q^{*\mu} \, \frac{t \cdot \epsilon_p^*}{t \cdot p}
+\frac{t \cdot\epsilon_p^*}{t \cdot p}\, \frac{t \cdot\epsilon_q^*}{t\cdot q}\, \frac{(q-p)^\mu}{2}
\\+\, 
\epsilon_p^{*}\cdot \epsilon_q^{*}\frac{(p-q)^\mu}{2p\cdot q} 
-\frac{p \cdot \epsilon_q^{*}}{ p \cdot q}\epsilon_p^{*\mu} 
+\frac{q \cdot \epsilon_p^*}{q\cdot p}\epsilon_q^{*\mu}
\bigg] \cM^{c,\mu}_\text{orig} (p+q)
	\label{SplitggME}
\end{multline}
The first line in \Eq{SplitggME} comes from the Wilson lines and the second from the usual self-collinear splitting diagram.

Now the original amplitude must satisfy a Ward identity, which holds exactly even if $p+q$ is off-shell.
Thus, $(p+q) \cdot \cM^{c}_\text{orig}  =0$.   Using this constraint, it is easy to check that $\cM^{ab}_\text{split}$ satisfies the Ward identity exactly for both outgoing gluons
(by replacing $\epsilon_p^{*\mu} \to p^\mu$ or $\epsilon_q^{*\mu} \to q^\mu$ ). 
Thus we can square the amplitude and sum over all polarizations with the simple replacement
$\epsilon^{*\mu}\epsilon^\nu \to - g^{\mu\nu}$, and sum over colors. The most enhanced terms will scale like  $|\cM_\text{orig}|^2 \lambda^{-2}$ and we can drop anything subleading. 
Since $(p+q) \cdot \cM^{c}_\text{orig}  =0$ and $p$ and $q$ are collinear, we also have 
$p \cdot \cM_\text{orig} \lesssim  \lambda$ and $q \cdot \cM_\text{orig} \lesssim \lambda$. 
The only terms that remain at leading power are therefore
\begin{align}
& \sum_\text{pols., cols.}\big| \cM^{ab}_\text{split} \big|^2 
\LPeq-\frac{2g^2 C_A}{ p\cdot q} \big(\cM^{c,\mu}_\text{orig}\big)^\dg\cM^{c,\nu}_\text{orig}
\bigg[
g^{\mu\nu} \frac{t \cdot p}{t \cdot q} + g^{\mu\nu} \frac{t \cdot q}{t\cdot p} 
+ \frac{p^\mu q^\nu }{p\cdot q} 
 \bigg]
 \label{polsummedexp}
\end{align}


To simplify this expression, it is helpful decompose $p^\mu$ and $q^\mu$ into a component in the $(p+q)^\mu$ direction which will vanish upon contraction with $\cM_\text{orig}$ by the Ward identity and a component orthogonal to the $\vec p+ \vec q$ which we call $\vec p_\perp$.
To keep $p^\mu$ and $q^\mu$ lightlike, their energies must be shifted slightly. 
The decomposition can be written as
\be
 p^\mu = z (p+q)^\mu +  p_\perp^\mu + \delta E\, (1,\vec0\,)^\mu
\And
q^\mu = (1-z) (p+q)^\mu -  p_\perp^\mu - \delta E\, (1,\vec0\,)^\mu
\ee
where
\be
\delta E = \frac{(1-2z)(p+q)^2}{2Q} \sim \lambda^2 
\ee
with  $Q=p^0+q^0$ and $p_T^2 = - p_\perp \cdot  p_\perp > 0$ is given by
\be
p_T^2 = z (1-z)(p+q)^2  + \delta E^2 \sim \lambda^2 
\label{pqscale}
\ee
At leading power, we can invert this last equation to write $p\cdot q = \frac{p_T^2}{2z(1-z)}+ \cO(\lambda^4)$.

Now we can simplify  \Eq{polsummedexp}. Since $t$ is not collinear to $p$ or $q$,
we have
\be
\frac{t\cdot p}{t\cdot q} \LPeq \frac{z}{1-z}
\label{tptqscale}
\ee
Next, we observe that the $p^\mu q^\nu$ term in \Eq{polsummedexp}  can be written as
$-p_\perp^\mu p_\perp^\nu$ at leading power because the Ward identity kills the $(p+q)^\mu$ terms. 
Thus,
\be
 \sum_\text{pols., cols.}\big| \cM^{ab}_\text{split} \big|^2 
\LPeq-\frac{2g^2 C_A}{ p\cdot q} \big(\cM^{c,\mu}_\text{orig}\big)^\dg\cM^{c,\nu}_\text{orig}
\bigg[
 g^{\mu\nu} \left( \frac{z}{1-z} + \frac{1-z}{z} \right) - 2z(1-z) \frac{p^\mu_\perp p^\nu_\perp}{p_T^2}
 \bigg]
\ee
which agrees with the polarized splitting function (cf. $\hat{P}_{gg}^{\mu\nu}$ in~\cite{Catani:1999ss}).
For the unpolarized splitting we discard spin correlations, by performing an average over azimuthal angle. 
This amounts to replacing in  \Eq{polsummedexp} 
\be
p^\mu q^\nu \to 
- p_\perp^\mu p_\perp^\nu \to 
-\frac{1}{2}\, p_T^2\, \delta_\perp^{\mu\nu} \to 
\frac12\, p_T^2 \,g^{\mu\nu}
\label{kperp2}
\ee
where each step is valid at leading power and the first and last arrows exploit the 
Ward identity on $\cM_{\text{orig}}$.

Inserting Eqs. \eqref{pqscale}, \eqref{tptqscale} and  \eqref{kperp2} into  \Eq{polsummedexp} we
then find
\begin{align}
\sum_\text{pols., cols.}\big| \cM^{ab}_\text{split} \big|^2 \LPeq \frac{2g^2 C_A}{p\cdot q}
\bigg[ \frac{z}{1-z} + \frac{1-z}{z} + z(1-z) \bigg] \times
 \sum_\text{pols., cols.} \big|\epsilon^*_{\rho} \cM^{c,\rho}_\text{orig}\big|^2
\end{align}
where the last factor is exactly the probability without splitting of the completely general process. The rest is the gluon-gluon splitting function in QCD.

In summary, we have shown that the quark-gluon and gluon-gluon splitting functions are universal and  reference-vector independent.


\section{Conclusions}
\label{sec:conclusions}
The main result of this paper is a proof at tree-level of factorization for matrix elements of operators in QCD. We show that matrix elements of operators with
$N$ fields in states whose momentum is either collinear to one of $N$ directions or soft can be written in a factorized form as
 \begin{equation}
\bra{X_1\cdots X_N; X_s} \bar\psi  \cdots \psi
\ket{0}
\LPeq
\bra{X_1} \bar{\psi} W_1 \ket{0}\cdots
\bra{X_N} W_N^\dag \psi \ket{0}
\bra{X_s} Y_1^\dg  \cdots Y_N\ket{0}
\label{overviewconc}
\end{equation} 
where $\LPeq$ means the two sides are equivalent at leading power in an expansion parameter $\lambda$ determined by the scaling of the momenta in the state $\bra{X_1\cdots X_N; X_s}$.
This equation with explicit color indices and gluon jets included is given in Eq.\eqref{genfactQCD}. In this equation both sides contain matrix elements of operators in QCD. That is, these are not effective field theory fields, although the connection to Soft-Collinear Effective Theory becomes trivial once this form is written down (see Section~\ref{sec:SCET}).

In order to prove \Eq{overviewconc} using only the scaling of momenta and not assigning scaling behavior to unphysical fields, we made critical use of the spinor-helicity formalism. Spinor-helicity methods let us assign scaling behavior to polarization vectors based on their momenta and the choice of an arbitrary reference vector. Crucially, the reference vector can be chosen differently for different gluons. By showing elements of factorization for certain reference vector choices and then showing reference-vector independence of the factorized result, the final factorization formula followed.

 Although choosing reference vectors sounds similar to choosing a gauge, the two are vastly different. Gauge choices are made for unphysical fields which can create and destroy any gluon state.
Thus one cannot assign different gauges to different sectors without chopping up the gauge field in some way, as in the effective field theory approach, or by attempting to formulate an incredibly nonlocal gauge condition. The spinor-helicity approach gets around awkward gauge conditions by choosing a reference-vector basis for the states directly, with the fields remaining in Feynman gauge (or whatever gauge one wants).

 We have proven \Eq{overviewconc} and its generalizations only at tree level.
 However, the equivalence probably holds to all orders in perturbation theory. The only modification should be that the right-hand side must be multiplied by a finite hard function $C(P_i)$ depending on the jet directions and energies but independent of $\lambda$. One can easily envision an all-orders proof which builds on the tree-level result, which contains all the infrared-singular real-emission graphs, and unitarity to relate the real-emission and virtual graphs. Indeed, unitarity constraints are efficiently encoded with on-shell methods like those we have employed here at tree-level. Pursuing this direction could conceivably lead to rigorous proofs of factorization for a wide variety of processes.

On a more practical side, a clean formulation of factorization, as in \Eq{overviewconc}, may lead to new calculations in perturbative QCD. For example, a
similar formula has already lead to one of the first studies of a jet-shape observable in SCET at subleading power~\cite{Freedman:2013vya}. Although so far, no results new to perturbative
QCD have been obtained this way, it is easy to imagine that subleading power factorization may eventually play a role in collider physics, as it has in heavy quark physics.


\section{Acknowledgements}
\label{sec:ackn}
We would like to thank John Collis, Davison Soper, George Sterman and Marat Freystis for helpful discussions.
The authors are supported in part by grant DE-SC003916 from the department of energy.
I.F. is also supported in part by the Natural Sciences and Engineering Research Council of Canada and the Center for the Fundamental Laws of Nature at Harvard University.

\bibliography{SoftCollPhys}
\bibliographystyle{h-physrev4}

\end{document}